\input{aipcheck}

\documentclass[
    ,final            
  ]
  {aipproc}

\layoutstyle{6x9}

\usepackage{latexsym}
\usepackage[english]{babel}
\usepackage[centertags]{amsmath}
\usepackage{amsfonts}
\usepackage{amssymb}
\usepackage{amsthm}
\usepackage{newlfont}
\usepackage{fancyheadings}
\usepackage{layout}

\newcommand{\re}{\ref}
\newcommand{\be}{\begin{equation}}
\newcommand{\ee}{\end{equation}}
\newcommand{\la}{\label}
\newcommand{\ber}{\begin{eqnarray}}
\newcommand{\eer}{\end{eqnarray}}

\newcommand{\bra}{\langle}
\newcommand{\ket}{\rangle}

\begin{document}

\title{PHYSICS OF ELECTROWEAK INTERACTIONS WITH NUCLEI}

\classification{25.30.-c; 21.45.-v; 21.60.De; 21.45.Ff}
\keywords{Lecture Notes}

\author{Giuseppina Orlandini}{
  address={Dipartimento di Fisica, Universit\`{a} di Trento\\ and INFN
Gruppo Collegato di Trento\\ Via Sommarive 14, I-38100 Trento (Italy)\\
orlandin@science.unitn.it}
}

\begin{abstract}
In this series of lectures it is illustrated how one can study the 
strong dynamics of nuclei by means of the electroweak 
probe. In particular, the most important steps to derive the cross sections 
in first order perturbation theory are reviewed. In the derivation
the focus is put on the main ingredients entering the hadronic part (response
functions), i.e. the initial and final states of the system 
and the operators relevant for the reaction.
Emphasis is put on the electromagnetic interaction with few-nucleon systems. 
The Lorentz integral transform method to calculate the response functions  
ab initio is described. 
A few examples of the comparison between theoretical and experimental results
are shown. The dependence of the response functions on the nuclear interaction and in particular 
on three-body forces is emphasized.
\end{abstract}

\maketitle
Emphasis is put on the electromagnetic interaction.

\section{Introduction}

Electroweak (e.w.) probes are essential to study the structure and the dynamics of nuclei. 
They help to shed light on  the most fundamental nuclear physics issues. In particular they allow 
\begin{itemize}

\item to assess the relevant degrees of freedom (d.o.f.) describing a 
nucleus. Traditionally these d.o.f. are supposed to be baryons and mesons. They  are called ``effective'' d.o.f.
to distinguish them from the fundamental d.o.f. of the strong interaction, i.e. 
quarks and gluons. Only in certain conditions of energy and momentum transferred by the e.w. probe to the nucleus
they become relevant to interpret experimental results.
These traditional effective d.o.f. of nuclear physics can also be divided in {\it explicit} and {\it implicit}
ones. The former are those appearing  in the Hamiltonian explicitly, namely the protons and the neutrons,
the latter are  hidden in the potential. A classical example of the latter d.o.f. is the pion
{\it implicit} in the One-Pion-Exchange Potential (OPEP).

\item to assess the potential model. Nowadays various realistic nucleon nucleon (NN) potentials are  available,
that reproduce thousands of NN scattering data with very high precision. While they are all equivalent in describing
the strong NN reaction, they are not, in principle, in describing a nuclear system undergoing an e.w. reaction. 
Therefore, by means of an e.w. probe one hopes to be able to discriminate 
among them, and have a better information about their origin.
Moreover, since the nuclear potential has an {\it effective} nature operating between composite systems, it is in principle 
a many-body operator. An aspect of nuclear dynamics that has attracted a lot of interest in the last years is 
the importance of multi-nucleon forces and in particular of the three-nucleon force (3NF). Electroweak reactions
can shed some light on their importance and origin.

\item to help understanding, by comparing theory and experiment, the microscopic origin of typical many-body  phenomenologies,
like for example {\it collectivity, clusterizations} or typical {\it single particle} (mean field)
behaviors.
\end{itemize}
The e.w. probe is mediated by the photon $\gamma$ or by the $W^{\pm}, Z_0$ gauge 
bosons. The coupling constants are small: the electromagnetic one $\alpha$ is 1/137, and
the weak one is even suppressed by a factor $10^{-5}$, due to the large mass of 
the weak gauge bosons. Therefore in calculating e.w. cross sections with hadrons 
it is perfectly legitimate to use the Born approximation, i.e. the one vector-boson 
approximation. This has the practical consequence to allow to separate the known information about 
the e.w. interaction from the unknown strong one.

\bigskip 

The plan of these notes is the following. 

\noindent In the first  section  I will outline the main steps 
which bring to the derivation of the e.w. cross section with an hadron. 
Extensive derivations can be found in a series of both classical and more modern books and articles (see e.g.
~\cite{Bjorken64}-\cite{Amaro05}).
I will deal with the electron scattering cross section as an example, suggesting how the procedure is modified in other
lepton scattering cases. The photoabsorption cross section will be viewed  as a particular case of the 
electron scattering cross section.

\noindent The second section will be an {\it intermezzo} to recall the scope of  e.w. studies and to point out
the interesting parts of the whole formalism.

\noindent The third section will be devoted to one of the main ingredients of the cross section i.e. the four-current
operator and its connections to the potential.

\noindent In the fourth section a few considerations about the importance of ab initio approaches and of the study 
of few-body systems will be made, giving a short overview of the theoretical problems and how they are 
treated in the literature. 

\noindent The fifth section will be devoted to review the Lorentz integral transform (LIT) method~\cite{LIT_REPORT}, describing
also its practical implementation.

\noindent In the sixth section an overview of interesting applications of the LIT method to electromagnetic cross sections will be
given, concentrating in particular on what can be learned about the multi-nucleon nature of the nuclear 
force. 

\noindent Finally it will be concluded summarizing the main messages contained in these notes.

\section{Outline of the derivation of electroweak cross sections }

\subsection{General considerations}

\begin{figure} 
  \includegraphics[width=.3\textheight]{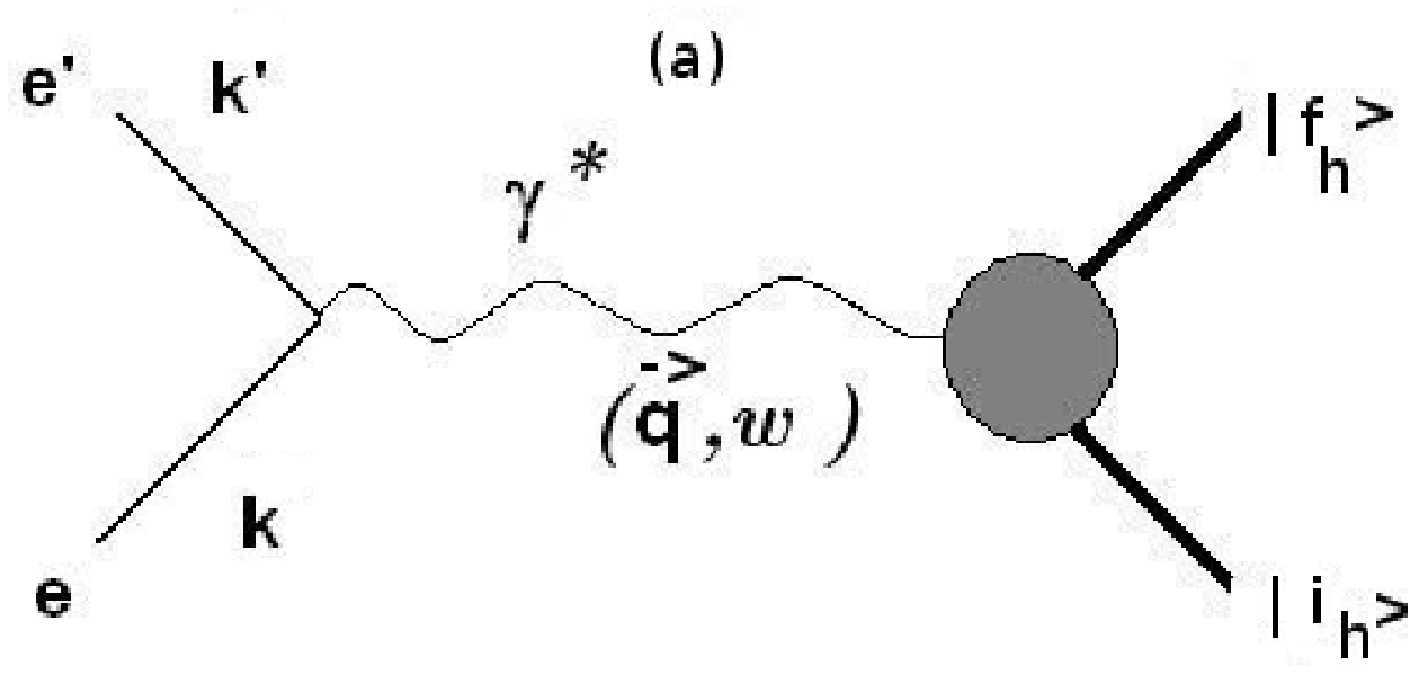}
  \includegraphics[width=.3\textheight]{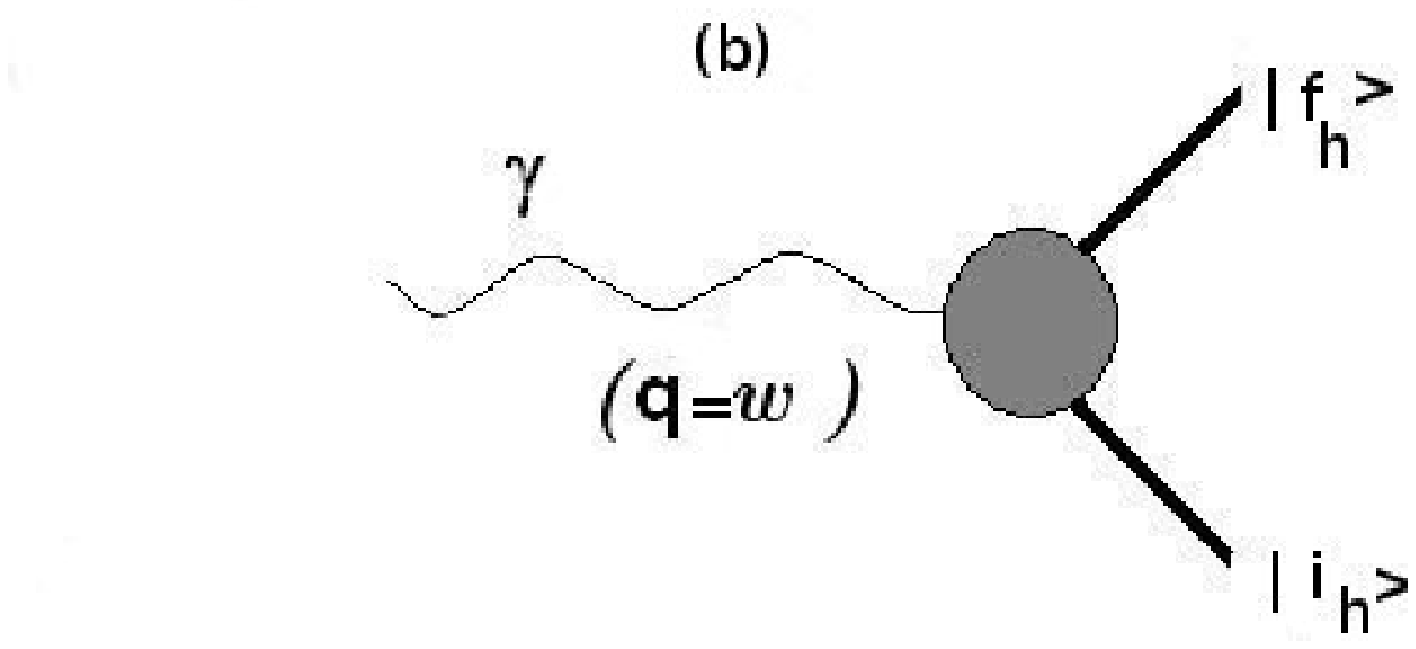}
  \caption{(a) Electron scattering in the one-photon exchange approximation, (b) photoabsorption.}
\label{figure1_ab}
\end{figure}

In this section I describe  how it is possible to
get information on the dynamics of a nucleus (and of an hadron system in general)
letting  it interacts
with an e.w. field and measuring the relative cross section. What I want to put in evidence in the
theoretical derivation of this cross section are the matrix elements $<f_h|j_h|i_h>$. Here
the initial and final state of the hadron are indicated by 
$|i_h>$ and $|f_h>$
respectively and $j_h$ is the four-current density operator (excitation
operator) which is responsible for the internal excitation of the target, via the interaction with the external field.
As was already mentioned in the introduction I specialize to the case of electron scattering, suggesting how the procedure is 
modified in other
lepton-scattering cases. The photoabsorption cross section will be viewed  as a particular case of the 
electron scattering cross section. Let's repeat that we work within the one photon exchange approximation, which is depicted 
in figure~\ref{figure1_ab}. In figure~\ref{figure1_ab}(a) $e$ and $e'$ represent the electron before and after the scattering, 
$k$ and $ k '$ are its initial and final four-momenta. Therefore one has:
\begin{equation}
 k=(E,\vec k)\,,\,\,\, k'=(E',\vec k')\,,\,\,\, E^2- |\vec k|^2=E'^2- |\vec k'|^2=  m_e^2\,.
\end{equation}
Momentum and energy transferred by the electron to the nucleus are indicated by $\vec q$ and $\omega$, respectively:
\begin{equation}
\vec q= \vec k-\vec k '\,,\,\,\,\omega = E-E'\,,\,\,\, \omega^2-|\vec q|^2\equiv q^2\equiv-Q^2\,. 
\end{equation}
In figure~\ref{figure1_ab}(a) $\gamma^*$ indicates the so called {\it virtual photon}. Its virtuality is given by the fact that $q^2$ (i.e. 
the square of the mass of this exchanged photon) is
not zero, and even negative. This means that $ Q^2$ is positive as it is demonsrarted in the following.
\begin{eqnarray*}
Q^2\,=\,{|\vec q|}^2\,-\,\omega^2\,&=&\,{\big(|\vec k|\,-\,|\vec k'|\,\big)}^2\,-\,{(E\,-\,E')}^2\cr
&=&\, {|\vec k|}^2\,+\,{|\vec k'|}^2\,-\,2\,\vec k\cdot\vec k'\,-\, E^2\,-\,{E'}^2\,+\,2\,E\,E'\cr
&=& {|\vec k|}^2\,+\,{|\vec k'|}^2\,-\,2\, |\vec k|\,|\vec k'|\,\cos{\theta}\,-\, 
{|\vec k|}^2\,-\,{|\vec k'|}^2\,+\,2\,|\vec k|\,|\vec k'|\cr
&=&\,2\, |\vec k|\,|\vec k'|\,(1\,-\,\cos{\theta})\cr
&=&\,4\,|\vec k|\,|\vec k'|\,{\sin^2{\frac{\theta}{2}}}\geq 0\,.
\end{eqnarray*}
Similar pictures as in figure~\ref{figure1_ab}(a) can be drawn for other 
lepton scattering processes. One has simply to replace the electron with the lepton of interest and the photon with the 
appropriate gauge boson. In the limit $Q^2=0$ the photon becomes real. Therefore the picture describing photoabsorption 
becomes figure~\ref{figure1_ab}(b).

The advantage of studying  an electron scattering process with respect to photoabsorption is the fact that one can vary energy 
and momentum transfer 
independently.  Therefore one may explore the cross section in  the $\omega|\vec q|$ plane. This aspect is
strictly connected with the possibility of investigating the dynamics of the nucleus not only at different excitations 
but also at different ranges, being even able to realize the composite internal structure of
nucleons. This happens at high momentum transfer when the wavelength associated to $|\vec q|$ is comparable or smaller than the nucleon size.  
The only restriction is the so called \emph{space-like} condition $Q^2\geq 0$.
Furthermore, in contrast to hadron probes, the virtual photon, as
the real one, has a much larger mean free path, so that it can
explore the whole target volume.
In fact hadron probes tend to interact only on the surface.\\
The large flexibility of the electrons with respect to hadrons and
real photons reflects on the structure of the cross section. As we will see
this contains longitudinal and  transverse components, which
allow one to have different detailed informations about
nuclear dynamics.

Our starting point is the differential cross section defined
as follows:
\begin{equation}
d\sigma\, =\, {\frac{|{\cal M}_{if}|^2}{V T|\vec j_i^e|\rho_{target}}}
\,\,{d\vec k'}\, {d\vec P_f}\,\rho_f\,,
\end{equation}
where $|{\cal M}_{if}|^2$ is the probability that the system goes from the initial to the final state
 (lowest order S-matrix) i.e. $|{\cal M}_{if}|^2=\frac{1}{2} \sum_{s,s'}
|<F|H_{int}|I>|^2$ with $|I>$ and $|F>$ the initial and final states
of the {\it whole} system (incident electron $+$ target);
$|\vec j_i^e|$ is the incident electron flux and the 
sum and average over the spin states of the initial and scattered electron $s$ and $s'$
mean that we
restrict to unpolarized electrons. The phase space density $\rho_f$ is connected to the differential momenta of the fragments:
\begin{itemize}
\item if the hadron remains intact \, $\rho_f = 1$
\item if the hadron breaks into 2 fragments 
$\,\Rightarrow \, d\vec P_1\,d \vec P_2\,  =\,  d\vec P_f\, d \vec p $\quad i.e.\, $\rho_f $\, =\, $ d\vec p $\,\,
\item if the hadron breaks into 3 fragments
$\,\, \Rightarrow \,\, d\vec P_1\,d\vec P_2\,d\vec P_3\, =\, d\vec P_f\, d \vec p_1\, d\vec p_2$\, etc., where  $\vec P_f$ is the final momentum of 
the hadron center of mass (c.m.), while the $\vec p_i$ indicate the relative momenta 
of the fragments.
\end{itemize}

In the next subsection we will elaborate on the matrix element $<F|H_{int}|I>$, however,
as already mentioned, what is important for us is the part 
which regards the target $<f_h|j_h|i_h>$, because the internal dynamics of the system  is contained in that. 
Therefore in following the evolution  of the formulas one has to keep the focus on that matrix element.
 
It is interesting to note that one can perform several different kinds of experiments with the electromagnetic probe.
One can make {\it elastic scattering} experiments where the nucleus is not excited and the energy transferred by the 
lepton is found exclusively 
in the target recoiling energy (of course this is never the case for photons). When this happens one has $|i_h>=|f_h>$.
 Therefore in such experiments
the  focus is on  bound state properties of the nucleus.
If the energy transfer serves in addition to excite the nucleus one speaks of {\it inelastic scattering}.
In this case the information contained in the cross section is not only on the bound state, but also on the excited states.
Depending on the energy transfer (and of the nucleus) they can be discrete states or continuum (break-up) states.  
One can also study absorption or  emission of photons. 

If the energies and momenta do not exceed a few hundred of MeV, 
and the nuclei
contain a not too large number of protons
a good  theoretical framework for calculating the cross sections 
of all these processes consists of first order (one-photon-exchange approximation) quantum electro-dynamics (QED) and 
non relativistic quantum mechanics.
For heavier nuclei the first order can become questionable. Since in the following I will 
concentrate mainly on
light nuclei this problem has no relevance.

\subsection{The transition matrix}

Let us now discuss the ingredients of the transition matrix  ${\cal M}_{if}=<F|H_{int}|I>$ whose square modulus
is the main term in the cross section.

The interaction Hamiltonian is
\begin{equation}
H_{int}=\int d^4 x j^{e}_\mu(x)\cdot A^\mu(x)\,,
\end{equation}
where $j^{e}_\mu(x)$ is the four-current density of the electron and
$A^\mu(x)$ is the electromagnetic four-potential created by the target,
\begin{equation}
A^\mu(x)=\int d^4y D_F(x-y) j_h^\mu(y)\,,
\end{equation}
where $D_F(x-y)$ is the Feynmann propagator. 

\noindent The incident electron current is 
\begin{equation}
j^{e}_\mu=-e \bar\psi(x)\gamma_\mu\psi(x)\,,
\end{equation}
where $\psi(x)$ is the Dirac spinor $\sqrt{\frac{m}{E V}} u(k,s) 
e^{-i k \cdot x}$.
Therefore one has:
\begin{eqnarray}
<F|H_{int}|I>&=&{- \int d^4\, x\,\, e m \over \sqrt{E E'} V} \bar u(k',s')\, \gamma_{\mu}\,  u(k,s)  
<e^{i k'\cdot x}|\int d^4\,y\, D_F(x-y)|e^{-i k \cdot x}>\cdot\cr
\nonumber \\
 &\cdot&<f_h|e^{i P_f\cdot y}\,j_h^\mu\,e^{-i P_i\cdot y}|i_h> {1 \over {(2\pi)^4} } \,.   
\end{eqnarray}
\noindent Performing the integrals 
in $d^4\,y$ and $d^4\,x$ one gets
\begin{equation}
<F|H_{int}|I>= {-e^2 m \over \sqrt{E E'} V} \bar u(k',s') \gamma_\mu u(k,s)  
\int {d^4 q\over q^2} \delta^{(4)}(k'-k+q)\delta^{(4)}(P_f-P_i-q) J^\mu\,.
\end{equation}

{\bf Notice:} It is here that the matrix element of interest $<f_h|j^\mu_h|i_h>$ appears. For economy 
of notations it has been denoted simply by $J^\mu$.

\noindent The integral in $d^4\, q$ gives
\begin{equation}
<F|H_{int}|I>= {-e^2 m \over \sqrt{E E'} V} \bar u(k',s') \gamma_\mu u(k,s)  
{\delta^{(4)}(P_f-P_i+k'-k)\over q^2} J^\mu\,,
\end{equation}
where $q=k-k'=P_f-P_i$.

\subsection*{The lepton and hadron tensors}
If one inserts this expression in the transition probability and use
the following relation~\cite{Bjorken64} 
\begin{equation}
[{(2\pi)}^4\, \delta^{(4)}\,(P_f-P_i\,+\,k'-k)]^2\,=\,\delta^{(4)}\,(P_f-P_i\,+\,k'-k)\cdot V\cdot T\, {(2\pi)}^4\,,
\end{equation}
one obtains
\begin{equation}
|{\cal M}_{if}|^2= {e^2 m^2 \over E E' V^2 }
{\delta^{(4)}(P_f-P_i+k'-k)\over q^4}\cdot V\cdot T \sum_{s'}
|\bar u(k',s')  \gamma_\mu u(k,s)|^2 
J^{\mu\star} J^\nu\,.
\end{equation}
Defining the {\bf lepton tensor}
\begin{equation}
w_{\mu\nu}\equiv
|\bar u(k',s')  \gamma_\mu u(k,s)|^2 
\end{equation}
and the {\bf hadron tensor} $W^{\mu\nu}\equiv J^{\mu\star} J^\nu$, and considering
that the electron flux can be written as $|\vec J_e|=\rho v_i={1\over V}
{E\over |\vec k| }$ 
the differential cross section becomes
\begin{equation}
d \sigma= {e^4 m^2 \over E' |\vec k|}{1\over q^4} \sum_{s'}w_{\mu\nu}W^{\mu\nu}
\delta^{(4)}\, d\vec k'\,d\vec P_f\,.
\end{equation}
Manipulating the lepton tensor by using the properties of the traces 
of Dirac matrices one has 
\begin{equation}
w_{\mu\nu}={2\over m^2}[k_\mu k_\nu'+k'_\mu k_\nu-g_{\mu\nu}
(k\cdot k'-m^2)] + i h \epsilon_{\mu\nu\alpha\beta}k^\alpha k'^\beta]\,,
\end{equation}
where $h$ is the helicity of the longitudinally polarized electron.
For unpolarized electrons such a contribution vanishes.
Let's concentrate just on the part of the cross section which does
not depend on the electron polarization, i.e.
\begin{equation}
d \sigma= {e^4 m^2 \over E' |\vec k|}{1\over q^4} {2\over m^2}
[k_\mu k_\nu'+k'_\mu k_\nu-g_{\mu\nu}(k\cdot k'-m^2)]J^{\mu\star} J^\nu
\delta^{(4)}\, d\vec k'\,\,d\vec P_f\,.
\end{equation}
\subsection{The use of charge conservation}

At this point one can make use of the continuity equation $q_\mu J^\mu=0$ (charge conservation) both for the electron 
and hadron currents. 
Remember that $d\sigma$ contains essentially ${|j^e_\mu J^\mu|}^2 $. The continuity equations imply 
$j^e_0={{\vec j^e \cdot \vec q} \over q_0}$ and $J_0={{\vec J \cdot \vec q}\over q_0}$. Therefore 
\begin{equation}
j^{e}_{\mu} J^{\mu} = j^{e}_{0} J^0 - \vec j_e \cdot \vec J = {{ (\vec q \cdot\vec  j_e) 
( \vec q \cdot J)} \over {q^2_0}} - \vec j_e \cdot \vec J = \vec j_e \cdot \Bigg( \vec q {{(\vec q \cdot \vec J)} 
\over {q^2_0}} - \vec J \Bigg) =  \vec j_e \cdot \vec J'\,,
\end{equation}
with $\vec J' \,=\, \vec J \,-\, \vec q {{( \vec q \cdot \vec J)}\over q_0^2}$.\quad 
This means that the use of the continuity equation 
allows to replace the 4-indexes $\mu,\nu=0,1,2,3$ with the 3-indexes
$i,j=1,2,3$ and the differential cross section becomes
\begin{equation}
d \sigma= {e^{4} \over {E' |\vec k|}}{2\over q^{4}}
\,\left [ k_i k'_j \,+\, k'_i k_j\, -\, g_{ij} \,(k \cdot k'- m^{2})\right ] \,J'^{i*}\, J'^{j}\,\delta^{(4)}\, 
d\vec k' \,\, d\vec P_f\,.
\end{equation}
Using the relation  \, $Q^2\,=\,-q^2\,=\,-\,(2 m^2-2 k\cdot k')$\, 
and performing the sum on $i,j$ one gets 
\begin{equation}
d \sigma= {e^4 \over E' |\vec k|}{2\over q^4}
\left [2 Re(\vec k\cdot \vec J'^*\,\vec k'\cdot \vec J')+{Q^2\over 2}
\vec J'^* \vec J\right ]
\delta^{(4)} \,d\vec k'\,\, d\vec P_f\label{dcross}\,.
\end{equation}

\subsection{Longitudinal and transverse {\it response} functions}
At this point it is convenient to choose a particular Cartesian system
where
\begin{equation}
\hat z=\hat q;\,\,\,\,\hat y={\vec k \times \vec k'\over|\vec k\times\vec k'|}
;\,\,\,\,\hat x=\vec y\times\vec z
\end{equation}
and decompose $\vec J'$ in
\begin{equation}
\vec J'=\vec J'_L+\vec J'_T\label{decompose}\,,
\end{equation}
with
\begin{eqnarray} 
\vec J'_L&=&\hat qJ'_L=(\vec J'\cdot \hat q)\,\hat q \cr
\vec J'_T&=& (\vec J'\cdot \hat x)\,\hat x+(\vec J'\cdot \hat y)\,\hat y\,.
\end{eqnarray}
Using the definition of $\vec J'$ and the continuity equation 
$J_0 q_0=\vec q\cdot \vec J$ one has 
\begin{equation}
J'_L=-{Q^2\over \omega |\vec q|} \,\rho\,.
\end{equation}
Substituting in~(\ref{dcross}) the target dynamics according to the decomposition~(\ref{decompose})
 one obtains three different contributions to the cross section i.e.
the longitudinal, transverse and mixed ones. Making use of a few kinematic
relations one has
\begin{equation}
(d \sigma)_L= {e^4 \over E' |\vec k|}{1\over q^4}{Q^4 \over |\vec q^4|} (4 E E'-Q^2)\, |\rho|^2 
 \delta^{(4)} \,d\vec k'\,\, d\vec P_f\,.
\end{equation}
Analogously the transverse and mixed contributions are given by 
\begin{eqnarray} 
(d \sigma)_T &=& {e^4 \over E' |\vec k|}{1\over q^4}4 |\vec k|\,|\vec k'| 
\left[\left({|\vec k|\,|\vec k'| sin^2\theta\over 2 |\vec q|^2}+
{Q^2\over 4 |\vec k|\,|\vec k'|}\right 
)\left(|J'_x|^2+|J'_y|^2\right)\right.+\cr
\nonumber \\
&+& \left.{|\vec k|\,|\vec k'| sin^2\theta\over 2 |\vec q|^2}
\left(|J'_x|^2-|J'_y|^2\right)\right] 
\delta^{(4)} \,d\vec k'\,\, d\vec P_f\,; 
\end{eqnarray}
\begin{equation}
(d \sigma)_{LT}= {-e^4 \over E' |\vec k|}{4\over q^4}{Q^2\over |\vec q|^3}
|\vec k|\,|\vec k'| sin\,\theta\, (E+E') Re(\rho^\star j'_x) \delta^{(4)} \,d\vec k'\,\, d\vec P_f\,. 
\end{equation}

The transverse contribution consists of  two separated terms which are
indicated with the labels $T$ and $TT$ respectively.
So at the end one can summarize the cross section as 
\begin{equation}
(d \sigma)= \left (V_L R_L +V_T R_T+V_{TT} R_{TT}+V_{LT} R_{LT}\right)
\delta^{(4)} \,d\vec k'\,\, d\vec P_f\,,\label{dsigma}
\end{equation}
where the V's indicate the coefficients of kinematic nature and the $R$'s the 
structure functions containing the matrix elements characterizing the target 
dynamics:
\begin{equation}
V_{L}= {e^4 \over E' |\vec k|}{1\over q^4}{Q^4 \over |\vec q^4|} (4 E E'-Q^2)\,;\label{VL}
\end{equation}
\begin{equation}
V_T={e^4 \over E' |\vec k|}{1\over q^4}\,2 |\vec k|\,|\vec k'| 
\left({|\vec k|\,|\vec k'| \sin^2\theta\over  |\vec q|^2}+
{Q^2\over 2 |\vec k|\,|\vec k'|}\right)\,;\label{VT}
\end{equation} 
\begin{equation}
V_{TT}= {e^4 \over E' |\vec k|}{1\over q^4} {2 |\vec k|\,|\vec k'| 
\sin^2\theta\over 2 |\vec q|^2}\,;\label{VTT}
\end{equation}
\begin{equation}
V_{LT}= {-e^4 \over E' |\vec k|}{4\over q^4}{Q^2\over |\vec q|^3}
|\vec k|\,|\vec k'| (E+E')\, \sin\,\theta\,; \label{VLT} 
\end{equation}
\begin{equation}
R_L=|\rho|^2\,;\label{RL}
\end{equation}
\begin{equation}
R_T=|J'_x|+ |J'_x|\,;\label{RT}
\end{equation}
\begin{equation}
R_{TT}=|J'_x|- |J'_x|\,;\label{RTT}
\end{equation}
\begin{equation}
R_{LT}= Re (\rho^\star J'_x)\,.\label{RLT}
\end{equation}

Equation~(\ref{dsigma}) is very general. The kinematic coefficients~(\ref{VL})-(\ref{VLT}) can be expressed in the non-relativistic 
or ultra-relativistic limits for the electron kinematics. 
In the ultra-relativistic limit i.e. when the mass of the electron is much smaller that 
its momentum  (that is certainly the case for electrons of tens or hundreds 
of MeV)  $E=|\vec k|$ and $E'=|\vec k'|$,
then one has
\begin{equation}
 {|\vec k'|}^2 V_{L}\, =\, \sigma_M {{ Q^4} \over {|\vec q|}^4}\,,
\end{equation}
where
\begin{equation}
\sigma_M\,=\,{1\over {4\,E^2}}\,{{{\cos^2{\theta\over 2}}}\over {\sin^4{\theta\over 2}}}\,\equiv \, 
\sigma_{Ruth.}\, {{\cos^2{\theta\over 2}}}\,;
\end{equation}
\begin{equation}
  {|\vec k'|}^2 V_T\,=\,{e^4 \over E' |\vec k|}\,{1\over q^4}\,2 |\vec k|\,|\vec k'| \, 
\left({|\vec k|\,|\vec k'| \sin^2\theta\over  |\vec q|^2}\,+\,
{Q^2\over 2 |\vec k|\,|\vec k'|}\,\right)\,;
\end{equation}
\begin{equation}
{|\vec k'|}^2 V_{TT}\,=\, {e^4 \over E' |\vec k|}{1\over q^4} {2 |\vec k|\,|\vec k'| 
\sin^2\theta\over 2 |\vec q|^2}\,;
\end{equation}
\begin{equation}
{|\vec k'|}^2 V_{LT}\,=\, {-e^4 \over E' |\vec k|}{4\over q^4}{Q^2\over |\vec q|^3}
|\vec k|\,|\vec k'| \sin\,\theta\, (E+E')\,. 
\end{equation}

\subsection{The cross section for inclusive, unpolarized electron scattering}

From   (\ref{dsigma}) one can obtain inclusive, seminclusive, exclusive, etc. cross sections according 
to what one is going to measure.\\
For example: if one wants the {\it inclusive} cross section i.e. one reveals the outcoming electron and nothing else
\begin{equation}
{d \sigma \over d \vec k'} = \int d \vec P_f \,(\, V_L R_L\,+\,V_T R_T\,+\,V_{TT} R_{TT}+V_{LT }R_{LT}\,) 
\,\delta^{(4)}\, (\, P_f\,-\,P_i\,+\,k'-k\, )\,.
\end{equation}
If one wants the {\it seminclusive} cross section i.e. one reveals the outcoming electron and a proton 
in coincidence \,$(\,\rho_f \,=\, d \vec p_P\,) $
\begin{equation}
{d \sigma \over {d \vec k' d \vec P_P}} = \int d \vec P_f\, (\, V_L R_L\,+\,V_T R_T\,+\,V_{TT} R_{TT}\,+\,V_{LT} R_{LT}\, )\, 
\delta^{(4)}\, (\, P_f\,-\, P_i\,+\, k'\,-\, k \,)\, {d \vec p_P \over {d \vec P_p}}\,,
\end{equation}
where $\vec p_P$ is the momentum of the proton in the c.m. system, while $\vec P_P$ is measured in the lab.
 system, etc.
Let us concentrate on the simplest  cross section i.e. the {\it inclusive} one. This is certainly the 
simplest from an experimental point of view, since 
one only needs (besides accelerator and target) only an electron spectrometer counting the electrons 
with a given energy, momentum at a certain scattering angle. 
A seminclusive cross section requires an additional hadron spectrometer and 
the number of counts in coincidence will be certainly smaller (smaller cross section).  From the
theoretical point of view, at a first sight,
the inclusive cross section  does not appear to be the simplest, but on the contrary in some cases the most 
complicate. I will comment about this in the section where  the LIT method is described.

A simplification of the expression for the cross section arises if one considers unpolarized targets. In fact 
in this case one has to sum and average on the target spin projections in the final and initial state, respectively.
One can show that this implies that
the terms $V_{TT} R_{TT}$ and $V_{LT }R_{LT}$ vanish. Therefore one has
\begin{eqnarray}
{ d \sigma \over d \vec k'}\, & = &\,  {d \sigma \over {d \Omega_e\, d |\vec k|'}}  \\
\nonumber
 & = & \,{ k'^2 \int d \vec P_f\, (\,V_L R_L\,+\,V_T R_T\,)\, \delta^{(3)}\, (\,\vec P_f \,-\, 
\vec P_i \,-\, \vec q\, )\, \delta\, (\, E^h_f\, -\, E^h_i \,-\, \omega\,) }\,, 
\end{eqnarray} 
where $ E^h_f\,,\, E^h_i$ represent the energies of the nucleus in the initial and final state. 
One has to stress here that these energies
are not only the internal energies, but include the recoil energy.
At this point one  substitutes  the matrix elements of interest in $ R_L$ and $R_T$ in 
(\ref{RL})-(\ref{RT}) and performs the integral in $d \vec P_f$, obtaining 
\begin{eqnarray}\nonumber
{d {\sigma} \over{d \Omega_e \,d\omega}}\,
& = & \,\sigma_M\,\left[
{ Q^4 \over {|\vec q|}^4} 
\,\sum_n\,{|\, \langle f_n\, |\, \widehat{\rho} \, |\,i_h \,\rangle|}^2\, \delta 
\,(E_f^h \,- \,E_i^h\, -\,\omega)\right.\\
&+&
\left.\left({ Q^2 \over {2 {| \vec q|}^2}}\, +\, {tg}^2 {\theta \over 2} \right) 
\,\sum_n\,{|\, \langle f_n\, |\, {\hat{\vec J}}_T \, |\,i_h \,\rangle|}^2\, \delta 
\,(E_f^h \,- \,E_i^h\, -\,\omega)\right]\,,
\end{eqnarray}
where the $\delta$-function $\delta \,(E_f^h \,- \,E_i^h\, -\,\omega)=\delta 
\,({\cal E}_f + E_{rec} - {\cal E}_i  -\,\omega) $ with ${\cal E}_f$  and ${\cal E}_i$
indicating the  internal energies and  $E_{rec} $ the recoil energy acquired by the nucleus 
(non relativistically $E_{rec}=
|\vec q|^2/2 M_A^2$ with $M_A$ representing the mass of the nucleus).

{\bf Notice:} The integral in $d \vec P_f$ has an important consequence: the operators 
$\widehat{\rho}$ and  ${\hat{\vec J}}_T$ are function of $|\vec q|$ and 
are expressed in the c.m. system of the nucleus. 

The following remarks are in order here.
\begin{itemize}
\item
 In order to get separate experimental information on $R_L$ and $R_T$ one needs to perform 
the so called {\it Rosenbluth separation} which consists 
essentially in performing two (or more) measurements of the cross section at fixed $\omega$ 
and $|\vec q|$ and different scattering angles. 
Representing these data in a XY plot where X = $V_T$ and Y is the cross section, one 
obtains a straight line whose properties are connected to 
$R_L$ and $R_T$.
\item
One can generalize the derivation of the cross section obtained above to take
into account also the {\it parity violating} 
contributions. This corresponds to add to the graph in figure~\ref{figure1_ab}(a)
a similar one with the $Z_0$ boson replacing the virtual photon.
\item
One can also generalize the procedure explained above to neutrino reactions ($\nu,\nu'$)  
($\nu,\mu^\pm$). In the former case  the virtual photon is replaced by 
$Z_0$. In the latter case by 
$W^\pm$. Of course in these cases the coupling constant $\alpha^2=e^2/\hbar c$ is replaced by the weak one.
\end{itemize}
\noindent 
One can read more extensive derivations of the e.w. cross sections in 
\cite{Donnelly85,Arenhoevel01,Amaro05}.

\section{Intermezzo}

\begin{figure}[htb] 
\centering
\includegraphics[angle=0,width=.50\textwidth]{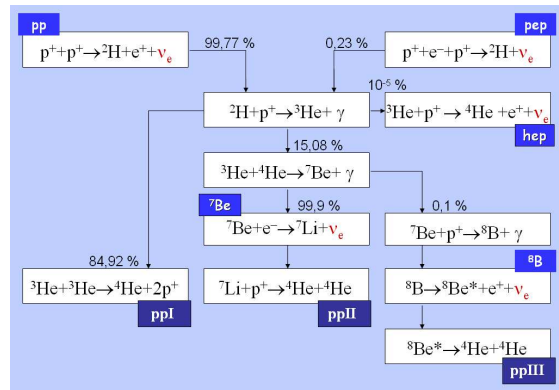}
\caption{The proton-proton cycles for the synthesis of $^4$He}
\label{figure2}
\end{figure}

\begin{figure}[htb] 
\centering
\includegraphics[angle=0,width=.50\textwidth]{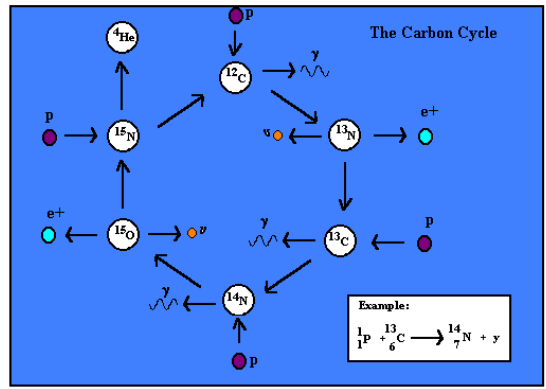}
\caption{The Carbon-Nitrogen-Oxygen cycle for the synthesis of $^4$He}
\label{figure3}
\end{figure}

At this point,  it is better to remind what is the scope of 
our study. In general one could distinguish two different attitudes in studying  e.w. interactions with nuclei.
One is what I will call the ``service'' attitude. It means that nuclear theorists  calculate e.w. cross sections
that are important to solve problems of astrophysical relevance. For example 
in figures~\ref{figure2} and~\ref{figure3}  one can see two famous nucleosynthesis cycles for the production of $^4$He. One can notice 
how many e.w. reactions are involved in the cycles. To know the different reaction cross sections is important 
to explain the abundances of 
elements in the Universe. Other e.w. cross sections are fundamental to explain the stellar evolution. Many of these cross sections
cannot be measured in the laboratory since it is often not possible to reproduce the astrophysical conditions.
Therefore nuclear theorists try to help to estimate those cross sections, as well as they can, using, in most cases, their experience with
models or experimental inputs from other sources or, in rarer cases, accurate ab initio calculations.  
The other attitude, that I would describe as a more ``fundamental'' attitude consists in trying to understand what are the 
fundamental degrees of freedom (implicit and explicit) at the nuclear scale; what are the properties of the nuclear force
and try to connect them to the underlying fundamental d.o.f. of QCD; what is the role of symmetries and their microscopic origin;
what is the microscopic origin of nuclear phenomena. All that requires ab initio approaches. It means that one
has to solve the many-body problem as accurately as possible, controlling the errors. Only in this case  the comparison 
of a theoretical result, obtained with a certain input, and the experimental data   gives information about how good 
the input is, without the uncertainties due to approximations in the solution of the many-body problem.

In deriving the inclusive unpolarized electron scattering cross section we have seen how the interesting and 
fundamental nuclear physics ingredients enter the cross section. They are contained in $|i_h \,\rangle$ and $\langle f_h\, |$
which are eigenfunctions of the nuclear Hamiltonian. Therefore they will be different for different nuclear potentials.
However, there is more information in the cross section: in fact the nuclear charge and nuclear current operators appear. 
This constitutes  an important difference between e.w. and purely hadron reactions. 
In the latter no information about charge and currents is present.
This is relevant, since it gives access to what one calls {\it underlying} or {\it implicit degrees of freedom}.
Let us explain this further in the next section.

\section{The nuclear four-current operator}

If one thinks that the most reasonable {\it effective} d.o.f. in a nuclear Hamiltonian are protons and neutrons it becomes
natural to describe  the interaction of the electron with the nucleus as a sum of the contributions of all the interactions 
with the nucleons, thought as Dirac particles with anomalous magnetic moments. 
However, since  one then calculates the matrix elements within non relativistic
quantum mechanics one has to perform a non relativistic reduction of this interaction.
This can be done via the so called  Foldy-Wouthousen transformation~\cite{FW50,EISGREI88}.
Essentially what one gets from such a  transformation  is an expression
for the charge and the current operators  at lowest relativistic order. The result is
very intuitive. For the density operator one obtains
\begin{equation}
\hat \rho(\vec q )= e \sum_{i=1}^Z \exp{(i \vec q \cdot \vec r_i)}\,.
\end{equation}
This corresponds to the Fourier transform of the sum  of Z $\delta$-functions centered in the positions
of the protons (as was already noticed above the positions have to be taken in the c.m. of the nucleus).
In this expression $e$ is the proton charge. The neutron has no charge so it does not appear. Of course
this ansatz is too drastic in that one knows that protons and neutrons have form factors. So what is done in practice is 
to replace $e$ with the proton form factor $G_E^p$ and sometimes even add the terms with the neutron 
contributions via $G_E^n$, even if in most cases they give negligible contributions.

At lowest order the expression for the transverse current is also intuitive. One gets two different currents, one due to the 
motion of protons, called {\it convection current} and the other due to the anomalous magnetic moments $\mu_i$ of both 
protons and neutrons, called {\it spin current};
\begin{equation}
\hat{\vec J}_c = e \sum_i^Z  \left\lbrace \frac{\vec p_i}{2m}, \exp{(i \vec q \cdot \vec r_i)} \right\rbrace 
\,,\label{JC}
\end{equation}   
\begin{equation}
\hat{\vec J}_s = \sum_i^A \mu_i \frac{\sigma_i \,\times\, \vec q}{2m}\exp{(i \vec q \cdot \vec r_i)}\,.\label{JS}
\end{equation}   
However, this is not all. In fact one has to notice that $\hat\rho,\hat{\vec J}$ and the Hamiltonian $\hat H$
are not independent. They have in fact to satisfy the continuity equation expressed by
\begin{equation}
  \vec\nabla \cdot \hat{\vec J} = -i [\hat H, \hat \rho]\,.\label{CC}
\end{equation}
So $\hat H$ and ${\hat \rho}$ fix the divergence of the current, but not the curl, therefore they do not fix $\hat{\vec J}$.
The convection and spin currents in (\ref{JC}) and~(\ref{JS})  are one body operators.
It is easy to prove that only the commutator $[\hat T, \hat\rho]$ actually reproduces the divergence of $\hat{\vec J}$, namely one has
\begin{equation}
\vec \nabla\cdot (\hat{\vec J}_c+\hat{\vec J}_s)= -i [\hat T, \hat \rho]\,.
\end{equation}
But then there must exist another current which is a two-(or many-)body operator which satisfies  
\begin{equation}
\vec\nabla\cdot \hat{\vec J}^{\,exc}= -i [\hat V, \hat \rho]\,.
\end{equation}
Therefore one has that a given potential can only fix the divergence of this current. But what about the curl?

If the potential is based on meson theory, one of course
knows the entire current. This is the current of the exchanged meson. Since this meson does not appear
as an explicit degrees of freedom in the Hamiltonian, it is an {\it implicit} d.o.f.. In this sense one says that the e.w.
interaction gets information on {\it implicit}  degrees of freedom, namely those intermediate ones,
 connecting nuclear
physics to the fundamental theory of strong interaction, i.e. QCD in the non perturbative regime. 

Contrary to potentials based on meson theories, phenomenological potentials are not built
knowing the  underlying degrees of freedom. Therefore such potentials 
may be very accurate in reproducing 
nucleon-nucleon  scattering data, but their reliability in electromagnetic interaction is
in principle unknown since the  {\it exchange currents} are not known.  
As already said charge conservation can give the constraint on the divergence of the currents, but 
no constraint on the curl exists. Therefore in order to calculate an electromagnetic cross section
one must have a ``model'' for the underlying degrees of freedom in the potential and  the
comparison with electromagnetic data will allow to judge its reliability.

\section{ab initio approaches and few-body physics}

As already explained above  {\it ab initio} calculations are those   requiring
the Hamiltonian $\hat H$, the four-current ($\hat \rho, \hat{\vec J}$) (provided the consistency in  (\ref{CC})) 
and the kinematic conditions of the reaction  as only
inputs, treating all degrees of freedom of the many--body system 
explicitly and accurately (microscopic approach). Only in this way the comparison theory-experiment can be
meaningful regarding the reliability of the inputs. 
However, ab initio approaches are a real challenge. At present   only   when the number of particles is  relatively small
 one is able to get accurate
solutions of the quantum mechanical many-body problem, without
the need of approximations. They are  necessary and unavoidable for more
complex systems. From here comes the importance of {\it few-body physics}.

Solving the quantum mechanical many-body problem has very different degrees of difficulty, depending 
if one deals with bound or continuum states. In general the situation is particularly problematic in nuclear physics,
due to the complicated structure of the nuclear potential (a brief discussion about the nuclear potential will be found later). 
In the following the problems of bound and continuum states are discussed separately.

\subsection{Bound states}\label{bound_states} 

Nowadays three- and four-nucleon bound states and binding energies can
be calculated with different methods based on any of the most modern 
high-precision NN-forces with an accuracy on the percentage level or 
less. For A=3 and 4 a well founded formulation, 
the Faddeev-Yakubovsky scheme (FY)~\cite{Yakubovsky,kamada,Gloeckletext},
opened that avenue followed by alternative, equally accurate procedures:
expansions in hyperspherical harmonics (HH)~\cite{A89,F83,VKR95} 
or gaussians (CRCGV)~\cite{Kamimura88,Kameyama89}, stochastic variational method
(SVM)~\cite{c1}, and path integral techniques in form of the 
``Green's Function Monte Carlo'' method (GFMC)~\cite{Carlson,AV18+,aequal8}.
Other very promising methods, based on the theory of effective 
interactions~\cite{LS80}, have been developed: the ``no-core shell model'' 
(NCSM)~\cite{Navratil99,Navratil00,Navratil00a} and the ``effective interaction HH''
(EIHH)~\cite{Barnea00,Barnea01} using expansions in harmonic oscillator and HH basis
functions, respectively.

\begin{table} 
\caption{Kinetic $\langle T \rangle$, potential $\langle V \rangle$,
binding energy $E_b$ (all in MeV),
and mean square radius of $^4$He as obtained by various
methods. From~\cite{benchmark4}}.
\begin{tabular}{lrrrrr}
  \hline  
  & \tablehead{1}{r}{b}{Method}
  & \tablehead{1}{r}{b}{$\langle T \rangle $}
  & \tablehead{1}{r}{b}{$\langle V \rangle $}
  & \tablehead{1}{r}{b}{$E_b$}    
  & \tablehead{1}{r}{b}{$\langle r^2\rangle^{1/2}$}   \\
  \hline
& FY     & 102.39     & -128.33      & -25.94(5)   & 1.485     \\
& CRCGV  & 102.25     & -128.13      & -25.89      &          \\
& SVM    & 102.35     & -128.27      & -25.92      & 1.486     \\
& HH     & 102.44     & -128.34      & -25.90(1)   & 1.483     \\
& GFMC   & 102.3(10)  & -128.25(10)  & -25.93(2)   & 1.490(5)  \\
& NCSM   & 103.35     & -129.45      & -25.80(20)  &           \\
& EIHH   & 100.8(9)   & -126.7(9)    & -25.944(10) & 1.486 \\
 \hline
\end{tabular} 
\label{table1}
\end{table}

An example of the degree of accuracy reached by these methods in the
four-body  system is given in table~\ref{table1}~\cite{benchmark4}, where 
binding energy, expectation values of kinetic and potential energies and 
mean square radius of $^4$He are shown as they result from calculations
of seven different  techniques based on the same potential model.
By the way, remembering that the experimental value of $^4$He is 28.3 MeV one can say that 
the most precise techniques lead to a clear cut answer: the potential used in~\cite{benchmark4}  underbinds
the $\alpha$-particle significantly. This is no accident.  It turns out that all modern
realistic  NN-forces underbind light nuclei significantly, showing the importance of three-body force (see below).
 \begin{figure}[htb] 
\includegraphics[angle=-90,width=.60\textwidth]{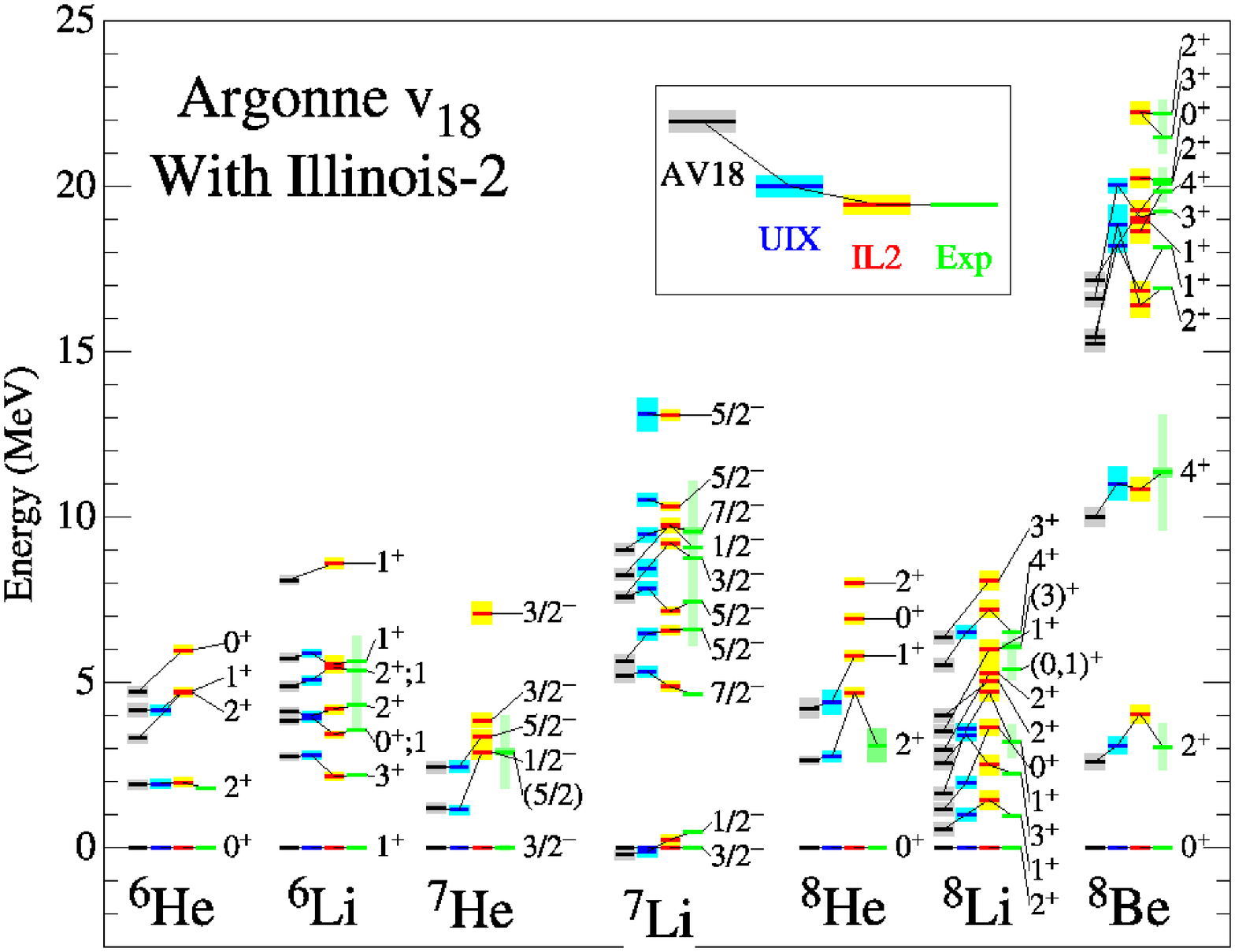}
\end{figure}
\begin{figure}[htb] 
\centering
\includegraphics[width=.60\textwidth]{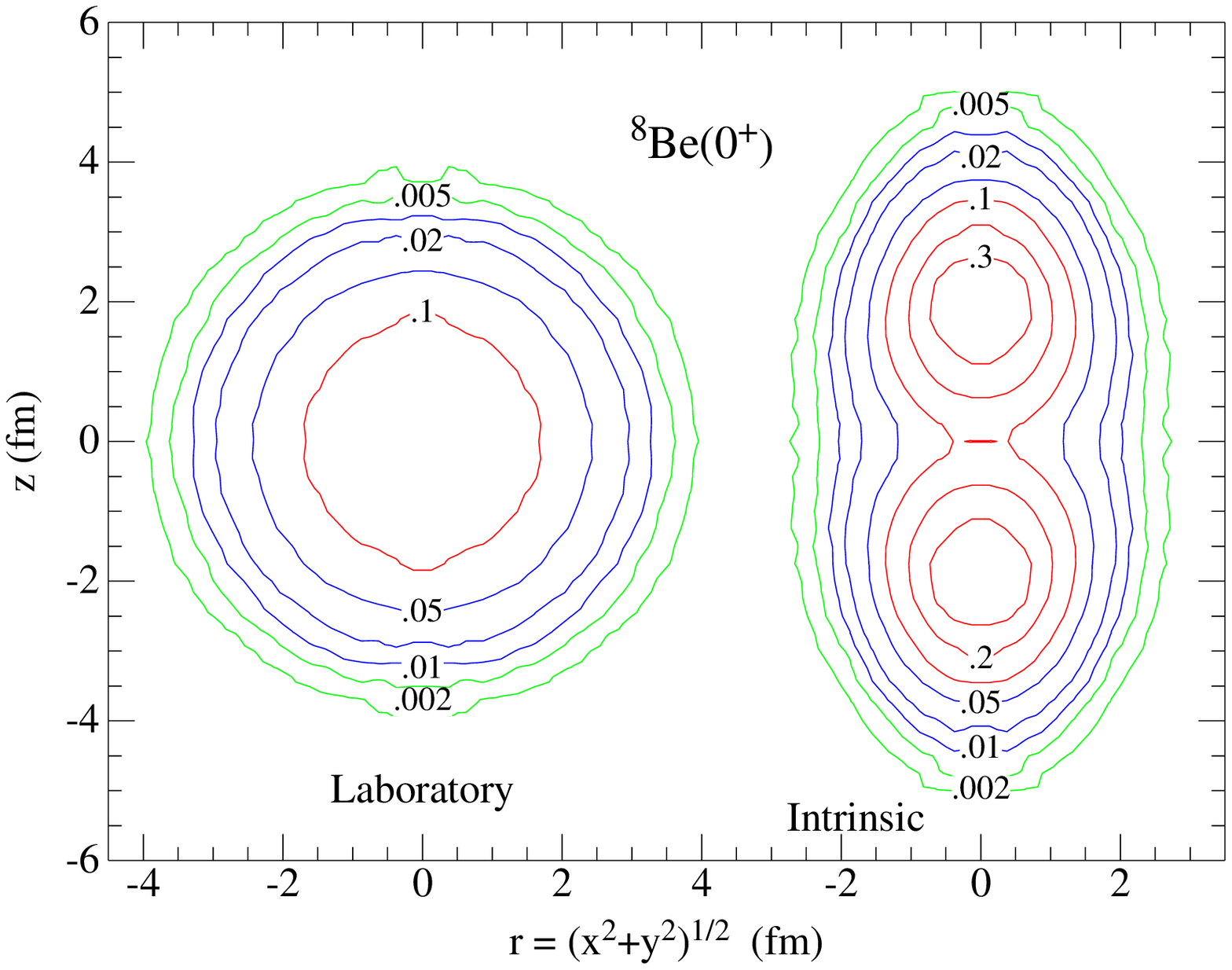}
\caption{
Upper panel: calculated spectra of A=6-8 nuclei from~\cite{wiringa00};
lower panel: calculated density contours of the $^8$Be ground state 
in the lab frame(left) and 
the intrinsic frame (right), 
labeled with densities in fm$^{-3}$.
}
\label{wiringa}
\end{figure}

While very light nuclei ($A=2-4$) 
are privileged systems for the study of fundamental issues like properties
and origin of the strong force, as is for example the existence of the multi-nucleon forces, 
the ab initio study of bound state properties of heavier nuclei 
has its own merit with respect to new many-body phenomena like, e.g.,
clusterization. Furthermore, one can expect that 
in heavier systems some of the interaction phenomena found in light nuclei
may be modified or amplified in view of the higher average nucleon density 
(``medium effects''). 

Going beyond four-body nuclei,
the low-lying spectra for up to A=8 nucleons
is rather well described~\cite{wiringa04}, as shown in table~\ref{table2}. 
Also these results clearly show that, in relation to the 
most modern NN-forces, 3N-forces are unavoidable in order to describe 
binding energies and low-lying excited states of light nuclei. 
Since the number of nucleon triplets overtakes more and more the number
of nucleon pairs with increasing A, it is clear that 3N-forces 
have to be included as well in any realistic description of complex nuclei.
\begin{table}
\caption{
Experimental and GFMC energies (in MeV) of particle-stable 
or narrow-width nuclear states. 
Monte Carlo statistical errors 
in the last digits are shown in parentheses. AV18 indicates the phenomenological NN potential~\cite{AV18}, while IL stands for 
a version of the Urbana 3NF. From~\cite{wiringa04}.
}
\begin{tabular}{lrrrr|lrrrrr}
\hline
  & \tablehead{1}{r}{b}{Nucleus}
  & \tablehead{1}{r}{b}{AV18}
  & \tablehead{1}{r}{b}{IL} 
  & \tablehead{1}{r}{b}{Exp} 
  & \tablehead{1}{r}{b}{Nucleus}
  & \tablehead{1}{r}{b}{AV18}
  & \tablehead{1}{r}{b}{IL}
  & \tablehead{1}{r}{b}{Exp}  \\
\hline
& $^3$H  ($\frac12^+$)  &  -7.61(1)  &  -8.44(1)  & -8.48 &
 $^7$Li ($\frac12^-$)  &  -31.1(2)  &  -39.0(2)  &  -38.77 \\
& $^3$He ($\frac12^+$)  &  -6.87(1)  &  -7.69(1)  &  -7.72 &
 $^7$Li ($\frac72^-$)  &  -26.4(1)  &  -34.5(2)  &  -34.61 \\
& $^4$He ($0^+$)        &  -24.07(4) &  -28.35(2) &  -28.30 &
 $^8$He ($0^+$)        &  -21.6(2)  &  -31.9(4)  &  -31.41 \\
& $^6$He ($0^+$)        &  -23.9(1)  &  -29.3(1)  &  -29.27 &
  $^8$Li ($2^+$)        &  -31.8(3)  &  -42.0(3)  &  -41.28 \\
& $^6$He ($2^+$)        &  -21.8(1)  &  -27.4(1)  &  -27.47 &
  $^8$Li ($1^+$)        &  -31.6(2)  &  -40.9(3)  &  -40.30 \\
& $^6$Li ($1^+$)        &  -26.9(1)  &  -32.0(1)  &  -31.99 &
  $^8$Li ($3^+$)        &  -28.9(2)  &  -39.3(3)  &  -39.02 \\
& $^6$Li ($3^+$)        &  -23.5(1)  &  -29.8(2)  &  -29.80 &
  $^8$Li ($4^+$)        &  -25.5(2)  &  -35.2(3)  &  -34.75 \\
& $^7$He ($\frac32^-$)  &  -21.2(2)  &  -29.3(3)  &  -28.82 &
  $^8$Be ($0^+$)        &  -45.6(3)  &  -56.5(3)  &  -56.50 \\
& $^7$Li ($\frac32^-$)  &  -31.6(1)  &  -39.5(2)  &  -39.24 &
  $^8$Be ($1^+$)        &  -30.9(3)  &  -38.8(3)  &  -38.35 \\
\hline
\end{tabular}
\label{table2}
\end{table}

For A~$=4,\, 6,\, 8,\dots$ 
one may already observe phenomena which are precursors of 
the above mentioned many-body phenomena. To illustrate this point, 
in the upper panel of figure~\ref{wiringa} we show the spectrum of $^8$Be as obtained in a 
microscopic ``Variational Monte Carlo'' - ``Greens Function Monte Carlo'' 
calculation (VMC-GFMC), based on one-body orbitals with four nucleons in an $\alpha$-core 
coupled to (A-4) one-body ($\ell=1$) wave functions.

\begin{figure}[htb] 
\includegraphics[width=.90\textwidth]{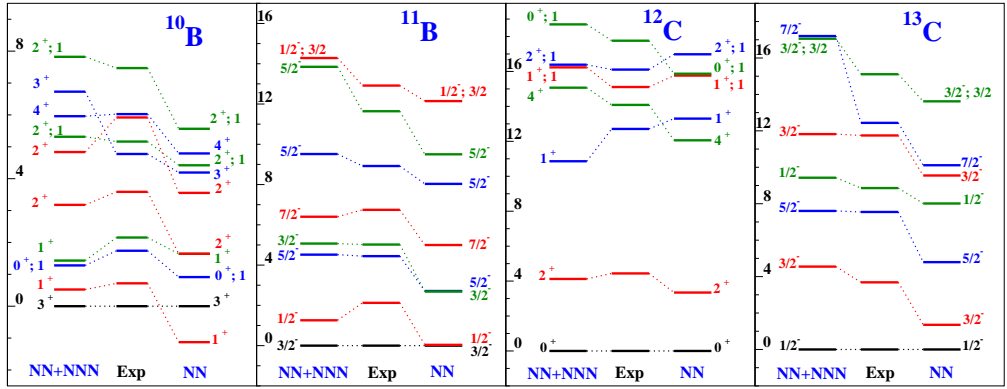} 
\caption{States dominated by p-shell configurations for $^{10}$B, $^{11}$B, $^{12}$C, and $^{13}$C. From~\cite{navratil09} }
\label{navratil} 
\end{figure}

Figure~\ref{navratil}  shows NCSM results~\cite{navratil09} for larger systems. Even if they are not fully converged they
show the power of the method. 
The Coupled Cluster approach~\cite{CCCoester,CCHagen09} is also very promising to explore the medium mass region. 

\subsection{Continuum states}

In general, if the reaction implies a state belonging to the continuum
spectrum of $\hat H$,  the challenge may become 
enormous, since one has to deal with the many--body scattering problem, which 
may lack a viable solution already for a very small 
number of constituents in the system. 
The difficulty in calculating a many--body cross section 
involving continuum states
can be realized, if one considers that, at a given energy,
the wave function of the system may have many different components (channels),
corresponding to all its partitions into fragments of various sizes.
Already in a rather small system of four constituents, at energies beyond the so--called
four--body break--up threshold, the two--, the three--
and the four--body break--up channels contribute. In configuration space the task consists in finding
the solution of the four--body Schr\"odinger equation with the proper
boundary conditions. It is just the implementation of the boundary conditions for 
a continuum wave function which constitutes the main obstacle to the practical solution
of the problem. In fact, the necessary matching of the wave function  to the 
oscillating asymptotic behavior (sometimes even difficult to be defined unambiguously)  
is not feasible in practice.
In momentum space the situation is as complicated. The proper extension of 
the Lippmann--Schwinger equation to a many--body system has been formulated long ago
with the  Faddeev--Yakubowski equations~\cite{FADDEEV:1961,YAKUBOWSKY:1967}. However,
because of the involved analytical structure of their kernels and the number of 
equations itself, to date it is impossible to solve the problem directly with their help,
at energies above the four--fragment break--up threshold, even for a number of constituents as 
small as just four.
\section{Integral transform methods}

Alternative approaches to the quite challenging problem of the dynamics
in the continuum are provided by integral transform methods.
The Lorentz Integral transform (LIT) method~\cite{EFROS:1994} is the natural extension of an original idea~\cite{EFROS:1985,LIT_REPORT}
to calculate reaction cross sections with the help of integral transforms. 
This kind of approach is rather unconventional. It starts from the consideration that 
the amount of information contained in the  wave function is  
redundant with respect to the transition matrix elements needed in the cross 
section. Therefore, one can avoid 
the difficult task of solving the Schr\"odinger
equation. Instead one can concentrate directly on the matrix elements. With the help
of theorems based on the closure property of the Hamiltonian eigenstates,
it is proved that these matrix elements (or some combinations of them) can be 
obtained by a calculation of an integral transform with a suitable kernel, 
and its subsequent inversion. The main point is that for some kernels the 
calculation of the transform
requires the solution of a Schr\"odinger--like equation with a source, and that 
its solutions have asymptotic conditions similar to a bound state. In this sense
one can say that the integral transform method reduces the {\it continuum} 
problem to a much less problematic {\it bound--state--like} problem.   
\begin{figure}[htb] 
\includegraphics[width=.70\textwidth]{F6a_example_R.eps}
\end{figure}
\begin{figure}[htb] 
\centering
\includegraphics[width=.70\textwidth]{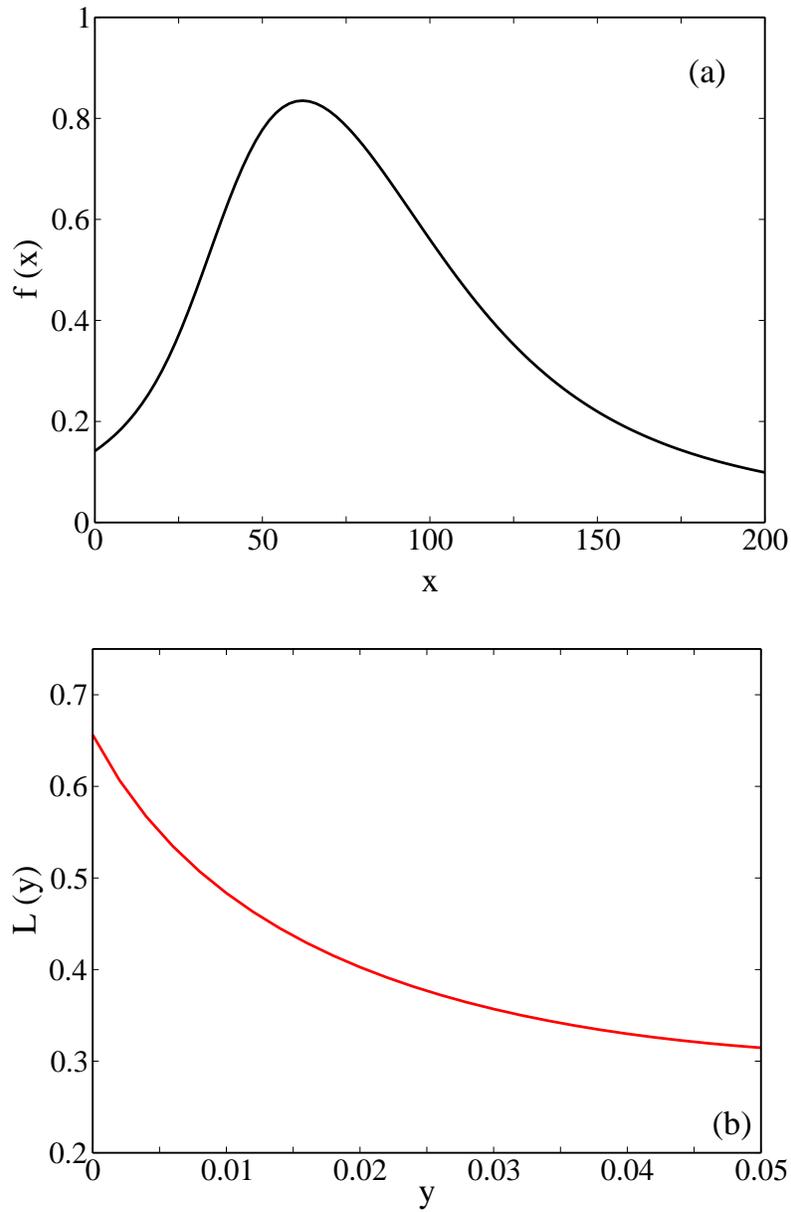}
\caption{
(a): the function $f(x)$;
(b): its Laplace transform $L(y)=\int f(x)\exp[- x y] dx$
}
\label{laplace}
\end{figure}
The form of the kernel in the integral transform is crucial. 
The reason is that in order to get 
the quantities of interest the transform needs to be inverted. 
Since it is normally calculated numerically it is affected by 
inaccuracies, and inverting an inaccurate transform is somewhat problematic.
Actually, when the inaccuracies in the
input transform tend to decrease, and a proper regularization is used in the course of inversion, 
the final result approaches the true one for various kernels~\cite{TIKHONOV:1977}. 
However, the quality of the result of the inversion may vary substantially according to
the form of the kernels, even for inaccuracies of similar size in the transforms.
In particular, when, for a specific kernel, the accuracy of the transform is 
insufficient, the result may be corrupted with oscillations superimposing the true solution.

In~\cite{EFROS:1985} the Stieltjes kernel was proposed and its reliability
was tested and discussed in simple model studies. Later, in a test of the 
method on a realistic electromagnetic cross section, calculated also in the conventional
way for the deuteron~\cite{EFROS:1993},
it was found that the use of the Stieltjes kernel is not satisfactory, since 
it leads to quite inaccurate results. The problem with the  Stieltjes kernel can 
be understood if one notices that its form is not qualitatively
different from that of the Laplace kernel. In fact it is well known that 
the problem of the inversion of a Laplace  
transform is extremely {\it ill posed} when the input is numerically noisy and incomplete
~\cite{ACTON:1970}. To illustrate the difficulties with the inversion of such kernels consider
figure~\ref{laplace}. In the upper and lower parts of the figure the function of interest and its 
Laplace transform are shown, respectively. As one can see the two curves do not resemble each other. The former
shows a curve with a bump, the latter a monotonically decreasing curve. If now one thinks that the 
second may be affected by numerical  errors as a result of the calculation one understands how difficult it
may be {\it to reconstruct} the curve in figure~\ref{laplace}(a). The Laplace kernel in fact has spread the information
about the curve over a large domain.  

Nevertheless the use of Laplace transforms 
is common in various fields of physics, from condensed matter to lattice QCD,
and  elaborated algorithms (e.g. the maximum entropy method~\cite{JAYNES:1978}) 
are sometimes employed for its inversion.

The problems encountered in inverting  the Stieltjes as well as the Laplace 
transform has led to the conclusion that, differently from those two cases,
the `best' kernel should be of a finite range. Its extension should be,  
roughly speaking,  about the range of the quantity to be obtained as the result 
of the inversion. Actually the perfect kernel would be a $\delta-$function.
In this case in fact the function and its transform would coincide. However, this 
clearly does not help! However, a so called {\it representation}
of the $\delta-$function would have the necessary characteristics.
 
At the same time, however, the transform has to be calculable in practice. 
In~\cite{EFROS:1994} it has been found
that the Lorentzian  kernel satisfies both requisites and the 
analogous test, as had been performed in~\cite{EFROS:1993} for the Stieltjes 
kernel, has led to very accurate results. 

\subsection{The Lorentz Integral transform  (LIT) method }

In the inclusive electron scattering process discussed in these lectures the quantity of interest
are the response functions $R_L$ and $R_T$ contained  in the cross section.
Like any general perturbation induced inclusive reaction (as are the e.w. reactions with hadron systems) they have the 
following structure
\begin{equation}
r(E)=
\sum\!\!\!\!\!\!\!\int\,d\gamma
\langle Q|\Psi_\gamma\rangle\langle\Psi_\gamma|Q'\rangle
\delta(E_\gamma-E)\,, \label{rr}
\end{equation}
where  $|\Psi_\gamma\rangle$ are solutions to the dynamic equation  
\be
(\hat{H}-E_\gamma)|\Psi_\gamma\rangle=0\,,
\ee
and $\hat{H}$ is the Hamiltonian of the system. 
The set $|\Psi_\gamma\rangle$  is assumed to be complete and orthonormal, 
\be  
\sum\!\!\!\!\!\!\!\int\,d\gamma|\Psi_\gamma\rangle\langle\Psi_\gamma|=1\,.\la{cl}
\ee
The integration and summation here and in (\re{rr}) go 
over all discrete states and continuum spectrum states in the set. 

We suppose that the norms $\langle Q|Q\rangle$ and $\langle Q'|Q'\rangle$ are finite. 
One has $|Q\rangle=\hat{O}|\Psi_0\rangle$, $|Q'\rangle=\hat{O}'|\Psi_0\rangle$, where 
$|\Psi_0\rangle$ is the initial state in a 
reaction (generally the ground state of the system undergoing the perturbation),
and $\hat{O}$, $\hat{O}'$ are transition operators. In the electron scattering case discussed above
they are both equal to the charge or to the current operators. Then one has
\begin{equation}
r(E)=
\sum\!\!\!\!\!\!\!\int\,d\gamma
\langle\Psi_0|\hat{O}^\dag|\Psi_\gamma\rangle
\langle\Psi_\gamma|\hat{O}'|\Psi_0\rangle
\delta(E_\gamma-E)\,. \label{re}
\end{equation}
Here  
$|\Psi_\gamma\rangle$ is a set of final states. 

As already pointed out, when the energy $E$ and the number of particles in the system increase
the direct calculation of the quantity $r(E)$
becomes prohibitive. The difficulty is related to the fact that in these cases
a great number of continuum spectrum
states $|\Psi_\gamma\rangle$ contribute to $r(E)$ and the
 structure of these  states is very complicated. 

The integral transform approach to  overcome this difficulty 
can be considered as a generalization of the sum rule approach,
since the use of the closure property of the Hamiltonian eigenstates plays a fundamental role.
Consider for example a simple sum rule for the quantity (\re{re}), based on the closure
property (\re{cl}) i.e.
\be    
\sum\!\!\!\!\!\!\!\int\,r(E)dE =\langle Q|Q'\rangle\,.\la{sum}
\ee
The  calculation of this quantity is much easier than a 
direct calculation of $r(E)$ itself,
since it requires the knowledge of $|Q\rangle$ and $|Q'\rangle$ only. This can be obtained 
with bound--state methods since it was  supposed that $|Q\rangle$ and $|Q'\rangle$ have finite norms.
However, this sum rule contains only a limited information on $r(E)$. In order to get
much more information about it we consider instead 
an integral transform
\be
\Phi(\sigma)=\sum\!\!\!\!\!\!\!\int\, K(\sigma,E)\,r(E)\,dE\la{phi}
\ee
with a smooth kernel $K$. This yields
\begin{eqnarray}
\Phi(\sigma)&=&\sum\!\!\!\!\!\!\!\int\,d\gamma
\langle Q|\Psi_\gamma\rangle K(\sigma,E_\gamma)
\langle\Psi_\gamma|Q'\rangle\nonumber\\
&=&\sum\!\!\!\!\!\!\!\int\,d\gamma
\langle Q|\hat{K}(\sigma,\hat{H})|\Psi_\gamma\rangle 
\langle\Psi_\gamma|Q'\rangle\,.\la{r}
\end{eqnarray}
Using the closure property (\re{cl}) one obtains
\be
\Phi(\sigma)=\langle Q|\hat{K}(\sigma,\hat{H})|Q'\rangle\,.\la{s}
\ee
Therefore equation (\re{s}) may be viewed as a generalized sum rule depending on 
a continuous parameter $\sigma$. 
Only with a proper choice of the kernel $K$ 
the right--hand side of (\re{s}) may be calculated using {\em bound--state} type
methods. And, as it was already said, even if $\Phi(\sigma)$ is available (\re{phi}) may not always  be solved with
enough accuracy to obtain $r(E)$ via an inversion of the transform. 

Our choice of the kernel $K(\sigma,E)$ is such 
that both the calculation of $\Phi(\sigma)$ and the inversion of (\re{phi}) are feasible. 
We choose~\cite{EFROS:1994}
\be
K(\sigma,E)=\frac{1}{(E-\sigma^*)(E-\sigma)}\,.\la{k}
\ee
Notice that the energy parameters $\sigma$ that we consider are complex.
For convenience we define them as
\be
\sigma=E_0+\sigma_R+i\sigma_I\,,\la{not}
\ee
where $E_0$ is the ground--state energy, and $\sigma_I\neq0$, so that
$K(\sigma,E)$ is actually a Lorentzian function centered on $E_0+\sigma_R$,
having  $\sigma_I$ as a half width
\begin{equation}
K(\sigma_R,\sigma_I,E)=\frac{1}{(E-E_0-\sigma_R)^2+\sigma_I^2}\,.\la{kRI}
\end{equation}
Then the integral transform (\re{phi}) becomes
\begin{equation}
L(\sigma_R,\sigma_I)=\sum\!\!\!\!\!\!\!\int\,dE\,
\frac{r(E)}{(E-E_0-\sigma_R)^2+\sigma_I^2}\,.\la{lorentz}
\end{equation}
Here and in the following the integral transform $\Phi(\sigma)$ with a Lorentz kernel 
is denoted by $L(\sigma_R,\sigma_I)$.
Using the definition (\re{re}) it is easy to show that the quantity (\re{lorentz}) may be  represented as 
\be
L(\sigma_R,\sigma_I)=\langle\tilde{\Psi}|\tilde{\Psi}'\rangle\,,\la{ov}
\ee
where the `LIT functions' $\tilde{\Psi}$ and $\tilde{\Psi}'$ are given by
\begin{eqnarray}
|\tilde{\Psi}\rangle&=&\left(\hat{H}-E_0-\sigma_R-i\sigma_I\right)^{-1}\hat{O}|\Psi_0\rangle\,, \la{psi1}\\
|\tilde{\Psi}'\rangle&=&\left(\hat{H}-E_0-\sigma_R-i\sigma_I\right)^{-1}\hat{O}'|\Psi_0\rangle\,.\la{psi2}
\end{eqnarray}
These functions are solutions to the inhomogeneous equations
\begin{eqnarray}
\left(\hat{H}-E_0-\sigma_R-i\sigma_I\right)|\tilde{\Psi}\rangle&=&\hat{O}|\Psi_0\rangle\,, \la{eq1}\\
\left(\hat{H}-E_0-\sigma_R-i\sigma_I\right)|\tilde{\Psi}'\rangle&=&\hat{O}'|\Psi_0\rangle\,. \la{eq2}
\end{eqnarray}
When $\hat{O'}=\hat{O}$, $L(\sigma)$
equals to $\langle\tilde{\Psi}|\tilde{\Psi}\rangle$. 
Since for $\sigma_I\ne0$
the integral in (\re{lorentz}) does exist, the norm of $|\tilde{\Psi}\rangle$ 
is finite. This implies
that $|\tilde{\Psi}\rangle$ is a {\em localized} function.
Consequently, (\re{eq1}) can be solved with bound--state type methods.
Similar to the problem 
of calculating a bound state it is sufficient to impose the only condition that the 
solutions  of~(\ref{eq1})  is localized. This means that
in contrast to continuum spectrum problems, in order 
to construct a solution, it is not necessary here to reproduce a complicated large 
distance asymptotic behavior in the coordinate representation or singularity structure 
in the momentum representation. This is a very substantial simplification.

Obviously, localized solutions to~(\re{eq1}) and~(\re{eq2})
are unique. Once $L(\sigma)$ is calculated  $r(E)$ is obtained by inversion of 
the integral transform with a Lorentzian kernel (\re{lorentz}) (`Lorentz integral transform'). 
The inversion of the LIT is discussed in the following.

\subsection{The inversion of the transform}

As already pointed out, the main advantage of the LIT method to study reactions
is that one avoids to solve the many--body scattering problem. One solves instead,
with bound--state methods, equations of the form (\ref{eq1}). The knowledge of those solutions 
leads to  the LIT  of the function $r$ of interest. A crucial part of the method 
is then the inversion of this integral transform. The inversion of this integral transform
has to be made with care, since errors in the transform can generate oscillations. To illustrate this
let us consider a well defined $r(E)$ to which we add a
high-frequency term $\Delta^\Omega r(E)$. The latter leads
to an additional $\Delta L^\Omega (\sigma_R,\sigma_I)$ in the transform. For any
amplitude of the oscillation $\Delta^\Omega L$ decreases with increasing $\Omega$.
This means that for some value of $\Omega$  $\Delta^\Omega L$ may be 
 smaller than the size of the errors in the
calculation. Therefore in this case  $\Delta^\Omega r$ cannot be discriminated. By reducing
the error in the calculation one can push the frequency of the undiscriminated $\Delta\Omega r$ 
to higher and higher values.

One of the methods that can be adopted to invert the transform is called the {\it regularization method}~\cite{TIKHONOV:1977}.
 This has led to very safe inversion
results. Alternative inversion methods are discussed in~\cite{ANDREASI:2005}. They
can be advantageous in case of response functions with more complex structures.
Up to now, however, such complex structures have not been encountered
in actual LIT applications, since
the various considered $r(e)$ have normally (i) a
rather simple structure, where essentially only a single peak of $r(e)$ has to be
resolved, or (ii) a more complicated structure, which however can be
subdivided into a sum of simply structured responses, where the various LITs can be 
inverted separately. 
The present `standard' LIT inversion method consists in the following ansatz for
the response function
\begin{equation}
r(e') = \sum_{n=1}^{N_{max}} c_n \chi_n(e',\alpha_i) \,,
\label{sumr}
\end{equation}
with $e'=e-e_{th}$, where $e_{th}$ is the threshold energy for the break--up
into the continuum. The $\chi_n$ are given functions with nonlinear parameters $\alpha_i$.
A basis set frequently used  for LIT inversions is
\begin{equation}
\label{bset}
\chi_n(\epsilon,\alpha_i) = \epsilon^{\alpha_1} \exp(- {\frac {\alpha_2 \epsilon} {n}}) \,.
\end{equation}
In addition also possible information on narrow levels
could be incorporated easily into the set $\chi_n$.
Substituting such an expansion into the right hand side of~(\ref{lorentz}) (here too 
the $\sigma_I$ dependence is omitted) one obtains
\begin{equation}
L(\sigma_R) =
\sum_{n=1}^{N_{max}} c_n \tilde\chi_n(\sigma_R,\alpha_i) \,,
\label{sumphi}
\end{equation}
where
\begin{equation}
\tilde\chi_n(\sigma_R,\alpha_i) =
\int_0^\infty de' {\frac {\chi_n(e',\alpha_i)} {(e'-\sigma_R)^2 + \sigma_I^2}}
\,\,.
\end{equation}
For given $\alpha_i$ the linear parameters $c_n$ are determined from a least--square best fit of
$L(\sigma_R)$ of equation~(\ref{sumphi}) to the calculated
$L(\sigma_R)$ of equation~(\ref{ov}) for a number of $\sigma_R$ points
much larger than $N_{max}$.

For every value of $N_{max}$ the overall best fit is 
selected and then the procedure is repeated for $N'_{max}=N_{max}+1$ till
a stability of the inverted response is obtained and taken as inversion
result. A further increase of $N_{max}$ will eventually reach a point, where the
inversion becomes unstable leading typically to random oscillations. The
reason is that $L(\sigma_R)$ of equation~(\ref{sumphi}) is not determined
precisely enough so that a randomly oscillating $r(e)$ leads to a better
fit than the true response. If the accuracy in the determination of $L(\sigma_R)$
from the dynamic equation is increased then one may include more basis functions in the expansion
(\ref{sumphi}).

The LIT method has to be understood as an approach with a {\it controlled resolution}. If
one expects that $r(E)$ has structures of  width $\gamma$ then the LIT resolution
parameter $\sigma_I$ should be similar in size. Then it is sufficient to determine the corresponding LIT 
with a moderately high precision, and the inversion should 
lead to reliable results for $r(E)$ if in fact no structures with a width smaller than $\Gamma$
are present. If, however, there is a reason to believe that $r(E)$ exhibits
such smaller structures one should
reduce $\sigma_I$ accordingly and perform again a calculation of $L$ with
the same relative precision as before. Such a calculation is of course more expensive than
the previous one with larger $\sigma_I$, but in principle one can reduce
the LIT resolution parameter $\sigma_I$ more and more. The advantage
of the LIT approach as compared with a conventional approach is evident. In the
LIT case one makes the calculation with the proper resolution, while in a
conventional calculation an infinite resolution (corresponding to $\sigma_I=0$) is requested, which
often makes such a calculation not feasible.

There are several tests of the reliability of the inversion. First of all, performing the calculation 
at different $\sigma_I$ one has to obtain the same stable result. If $\sigma_I$
is too small the solution tends very slowly to zero, therefore for $\sigma_I \geq \sigma_I^{min}$  one may have numerical problems, 
turning into large errors for the LIT. As already said above, this will show up in unphysical oscillations in $r(E)$.
This means that the stable result obtained with $\sigma_I < \sigma_I^{min}$ is the correct one, at that resolution. 
Another test is the control of the moments, in fact the moments of $r(E)$ can be calculating both integrating $r(E)$ or 
by expression~(\ref{s}) (with $K(\sigma, H) = H^\sigma$ and $\sigma$ integer), which needs  only the bound state.

\subsection{Practical calculation of the LIT}

We have seen how the LIT method reformulates a scattering
problem as a Schr\"odinger--like equation with source terms
which depend on the kind of reaction under consideration. These equations, 
which
we will call the `LIT equations` are essentially the same for any reaction, 
differing by the source 
term i.e. by their right hand side (see~(\ref{eq1}), (\ref{eq2})). 
In all cases the asymptotic boundary conditions are bound--state--like.
Consequently the solutions of these equations can be found with similar
methods as for the
bound--state wave functions. 

As we have seen above bound state solutions of the Schr\"odinger equation can be found in 
different ways searching for a direct numerical solution of the differential
equations in coordinate space (or of the integro--differential equations in 
momentum space), or alternatively using expansions
on some basis set of localized functions (bound-state-like boundary condition). 
These expansion methods become more and more  advantageous with respect
to the former one, with increasing particle number. For this reason, and because it has been used
for the vast majority of the LIT applications I discuss it in the following.
The truncation of the basis set converts the bound--state Schr\"odinger equation into a
matrix eigenvalue problem and the LIT equations into a set of linear equations.
These equations can be solved with various iteration methods
and also Gauss type non--iteration ones. Such strategies have the drawback that
one should solve these equations many times,
as many as the number of
$\sigma_R$ values that one needs for a proper inversion of the transform.
There are, however,  two better 
strategies for calculating
$L(\sigma)$. The first strategy, which is called the eigenvalue method involves the full diagonalization of the Hamiltonian 
matrix and expresses
the LIT through its eigenvalues. This method is instructive from a theoretical point 
of view.

Regardless of the reaction under consideration and the process
that one wants to study the LIT method requires the calculation of the overlap
(see~(\ref{ov})--(\ref{psi2}))
\begin{eqnarray}\label{L_ssp}
L(\sigma_R,\sigma_I)&=&\langle\tilde\Psi|\tilde\Psi'\rangle\nonumber\\
&=&\bra Q | \frac{1}{(\hat{H}-E_0-\sigma_R+i\sigma_I )}
           \frac{1}{(\hat{H}-E_0-\sigma_R-i\sigma_I )}
| Q'\ket\,,
\end{eqnarray}
where  $|Q'\ket$ and $|Q\ket$ contain the information
about the kind of reaction one is considering.
Seeking $|\tilde{\Psi}\rangle$ and $|\tilde{\Psi}'\rangle$ as expansions over $N$ localized basis states, 
it is convenient to 
choose as basis states $N$ linear combinations of states that diagonalize the Hamiltonian matrix. 
Let's denote these combinations $|\varphi_\nu^N\rangle$ and the eigenvalues $\epsilon_\nu^N$. 
The index $N$ is to remind that they  both depend on $N$. 
If the continuum starts at $E = E_{th}$ then at sufficiently high $N$ the states $|\varphi_N\rangle$ having 
$\epsilon_\nu^N< E_{th}$ will represent approximately the bound states. The other states  will 
gradually fill in the continuum as $N$ increases. The expansions of our localized  LIT functions 
read as 
\begin{equation}\label{nu1}
|\tilde{\Psi}\rangle =\sum_\nu^N\frac{\langle\varphi_\nu^N|Q\rangle}
{\epsilon_\nu^N-E_0-\sigma_R-i\sigma_I}|\varphi_\nu^N\rangle\,,
\end{equation}
\begin{equation}\label{nu2}
|\tilde{\Psi}'\rangle=\sum_\nu\frac{\langle\varphi_\nu^N|Q'\rangle}
{\epsilon_\nu^N-E_0-\sigma_R-i\sigma_I}|\varphi_\nu^N\rangle\,.
\end{equation}	
Substituting  (\ref{nu1}) and (\ref{nu2}) into (\ref{L_ssp}) 
yields the following expression for the LIT,
\begin{equation}\label{L_epsnu}
L(\sigma_R,\sigma_I)=\sum_{\nu} \frac{\bra Q| \varphi_\nu^N \ket \bra \varphi_\nu^N | Q' \ket}
                    {(\epsilon_{\nu}^N-E_0-\sigma_R)^2+\sigma_I^2}  \;,
\end{equation}
and for $|Q '\ket=|Q \ket$
\begin{equation}\label{L_epsnu_diag}
L(\sigma_R,\sigma_I )=\sum_{\nu} \frac{ |\bra \varphi_\nu^N | Q \ket|^2}
                     {(\epsilon_{\nu}^N-E_0-\sigma_R)^2+\sigma_I^2}\;.
\end{equation}
From~(\ref{L_epsnu}) and (\ref{L_epsnu_diag}) it is clear that
$L(\sigma_R,\sigma_I)$ is a sum of Lorentzians. The inversion of the LIT contributions
from the states with $\epsilon_\nu^N < E_{th}$ gives the discrete part of a response function, 
whereas the inversion of the rest gives its continuum part. The spacing between 
the corresponding eigenvalues with $\epsilon_\nu^N > E_{th}$ depends on $N$ and in a given energy 
region the density of these eigenvalues increases with $N$. 
(Since the extension of the basis states grows with $N$, this resembles the increase 
of the density of states in a box, when its size increases.) 

The second strategy to obtain the LIT utilizes the Lanczos algorithm~\cite{LANCZOS:1950,GOLUB:1983}
expressing  the LIT as a continuous fraction. 
It turns out that for obtaining an accurate LIT only a
relatively small number of Lanczos steps are needed.
As the number of particles in the system under consideration increases the
number of basis states grows up very rapidly  and the Lanczos approach seems at present 
the only viable method to calculate the LIT.

The motivation to use the Lanczos approach comes from the observation that from the computational 
point of view the calculation of the LIT is much more 
complicated and demanding than finding the ground state wave function of 
an $A$--particle system.
In fact, to  obtain the ground--state wave function one needs only to find the lowest
eigenvector of the Hamiltonian matrix. On the contrary, as it is clear from~(\ref{L_epsnu}) 
and (\ref{L_epsnu_diag}), 
the complete spectra of $\hat H$ over a wide energy range should be known
to calculate the LIT. Therefore it is no surprise that the
computational time and the memory needed to calculate $L(\sigma)$ have been the
limiting factors in extending the LIT method for systems with more than four
particles.
It turns out that these obstacles can be overcome if 
the LIT method is reformulated using the Lanczos algorithm.
Following reference~\cite{MARCHISIO:2003}, in this section it is shown
how this can be done. 

To this end it is assumed that the source states $|Q \ket$ and $|Q '\ket$ are
real and 
rewrite the LIT in the following form
\begin{equation} \label{lorelan1}
 L(\sigma)=-\frac{1}{\sigma_I} \mbox{Im} 
\left\{ \bra Q | \frac{1}{\sigma_R+i\sigma_I+E_0-
\hat{H}} | Q'\ket \right\}\:\mbox{.}
\end{equation}
A similar relation connects the response function $r(e)$ to 
the Green's function
\begin{equation} \label{lorelan2}
  r(e)=-\frac{1}{\pi} \mbox{Im} \left\{\lim_{\eta\to 0} 
  G(e + i \eta + E_0)\right\}
\:;\hspace{6mm} 
  G(z)=\bra Q | \frac{1}{z-\hat{H}} | Q' \ket \:,
\end{equation}  
provided that $z = e + i \eta$ is replaced by $\sigma_R + i \sigma_I$.
This is not surprising since the properly
normalized Lorentzian kernel
is one of the representations of the  $\delta$--function and $\sigma_I/\pi\,L(\sigma_R)  
\rightarrow r(\sigma_R)$ for $\sigma_I\rightarrow 0$. In condensed 
matter calculations~\cite{DAGOTTO:1994,HALLBERG:1995,KUEHNER:1995} the Lanczos
algorithm  has
been applied to the calculation of the Green function with a small value of 
$\eta$, and its imaginary part has been interpreted as $r(e)$
directly. This can be done if the spectrum is discrete (or discretized) and
$\eta$ is sufficiently small. In our case we have a genuine continuum
problem and we want to avoid any discretization, therefore we calculate
$L(\sigma_R)$ in the same way, i.e. with finite $\sigma_I$ using the 
Lanczos algorithm, but then we anti-transform $L(\sigma_R)$ in order to 
obtain $r(e)$.

\section{application of the LIT method to electroweak response functions}

In this section I present some selected results obtained with the LIT method for both 
the electron scattering response functions and the photoabsorption cross section.
In the various cases the LIT equation~(\ref{eq1}) has been solved with different 
ab initio bound state methods. For A$\geq 4$ the most efficient one has turned out to be the
EIHH method~\cite{Barnea00,Barnea01}, which, till now, has allowed to reach results up to A=7. 

The selection of reselts presented in the following is done  with two scopes: the first is to illustrate how
from the comparison between theoretical ab initio results and experimental data one can
get information on the nuclear force. The second is to show how, when A increases,  a microscopic calculation
can give rise to macroscopic features, giving the possibility to study the link between
the two scales (see the case A=6).

However, before entering into that discussion it is worth showing how the LIT method
is working in systems where the direct calculation of  continuum states is possible.
This is the case of the two-body system, i.e. the deuteron.  

\subsection{Results for A=2}

The very first LIT application was the calculation of
the deuteron longitudinal form factor $R_L(q,\omega)$ at $|\vec q|=440$~MeV/c in ~\cite{EFROS:1994},
where the LIT has been originally proposed. Actually it served  as a test case for the applicability of 
the LIT approach, since one can calculate $np$ continuum state wave functions explicitly. There it was shown 
how good was the agreement between the responses obtained in the two ways. Here another example 
on the total deuteron photodisintegration is shown. 
\begin{figure}[htb] 
\centering
\includegraphics[width=0.8\textwidth]{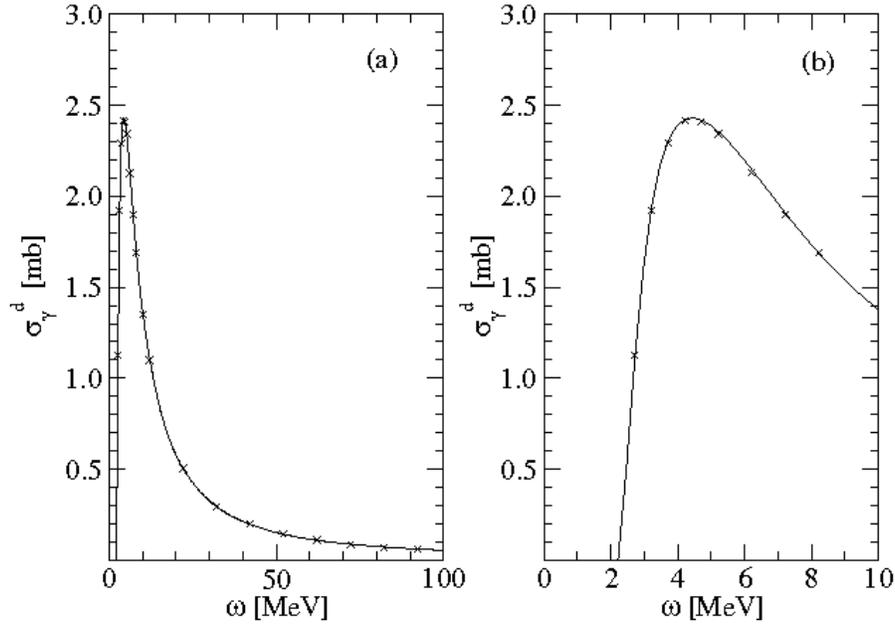}
\caption{Total deuteron photoabsorption cross section up to 100 MeV (a) and 10
MeV (b): LIT result (solid) and from calculation with explicit np final state wave
function (crosses).}
\label{testdeut}
\end{figure}

Figure~\ref{testdeut} shows the comparison of the LIT result with that of the conventional calculation,  using
the proper np scattering states. Also here one observes an
excellent agreement between the two calculations, showing the high precision that can
be obtained with the LIT method.

At this point let us make a digression about the operator which is very often used in the photoabsorption calculations, i.e.  the 
dipole operator $D_z=\sum_i z_i \tau_i^3$. Let us briefly explain why one uses such an operator in the 
(low energy) photoabsorption cross section. 

The photoabsorption cross section  can be obtained 
 by the electron scattering one in the limit $Q^2=0$ ($|\vec q|=\omega$). This means that the longitudinal part of the cross section
disappears (the photon has no charge). Therefore one remains with the part where the transverse current operator is present
(the photon fields are transverse to $\vec q$). If one makes a multipole expansion of the current operator and works out 
the cumbersome formulas, at some point one realizes that there are terms that contain the $\vec \nabla\cdot\hat{\vec J}$. 
This is nice because, 
at least for these terms, it is possible to use the continuity equation  again, avoiding the knowledge of the current. 
So one can connect these terms to the charge density multipoles. The interesting  fact is that the remaining terms
are of higher order in  $|\vec q|$ and therefore can be neglected if  $|\vec q|$, (i.e. $\omega$) is small enough.
However, if  $\omega$ is small enough, one can also neglect all the multipoles higher than the first, i.e. one remains with the dipole.
This is what is called {\it Siegert Theorem}. The great importance of this theorem is the fact that one can avoid to
know  $ \hat{\vec J}$ ! The information about it is intrinsically contained in the use of the dipole operator.
For our scope this has a very important consequence: if we use a phenomenological potential in the calculation, 
the comparison theory-experiment will tell us  whether the unknown {\it implicit degrees of freedom}, implied by it, are the
right ones.

\subsection{Results for A=3}

\begin{figure} 
  \includegraphics[height=.4\textheight]{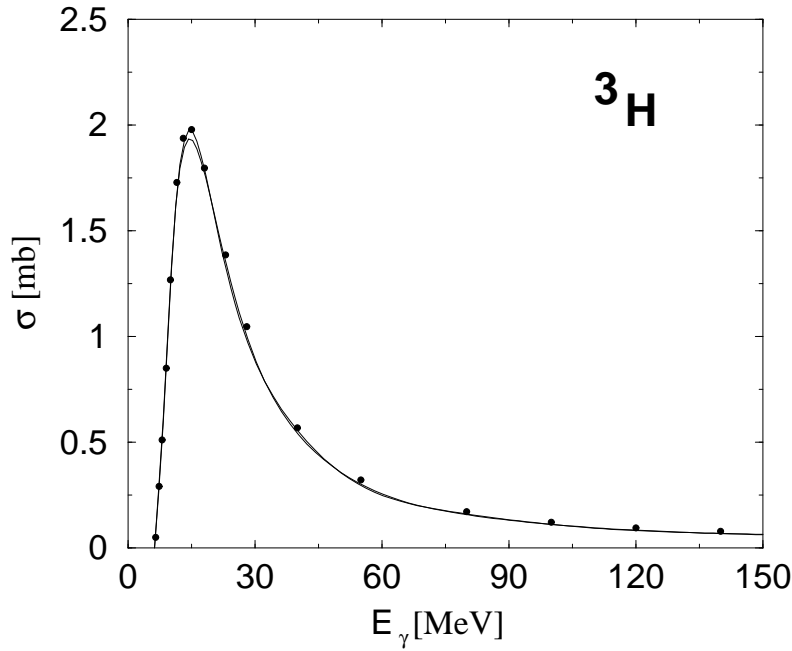}
  \caption{Comparison of the Faddeev and LIT results for the total $^3$H photoabsorption cross
section. The dots are the
Faddeev results and the two curves represent the bounds for the inversion of the LIT. From~\cite{bench3}}
\label{bench3}
\end{figure}

For the 3-body problem in the continuum one can  solve the Faddeev equations and calculate
continuum states both in the two- and three-body break up channels. Therefore also in this case one can benchmark traditional results
with those obtained by the LIT. This has been done in~\cite{martinelli,bench3}. Reference~\cite{martinelli}  is interesting 
because there the LIT equation
has been solved translating it  into the Faddeev scheme, showing once more that it can be solved with different techniques.
The excellent result of the benchmark from~\cite{bench3}  is reported in figure~\ref{bench3}.
Notice that in the figure the small uncertainty due to the inversion is also shown.

\begin{figure} 
  \includegraphics[height=.5\textheight]{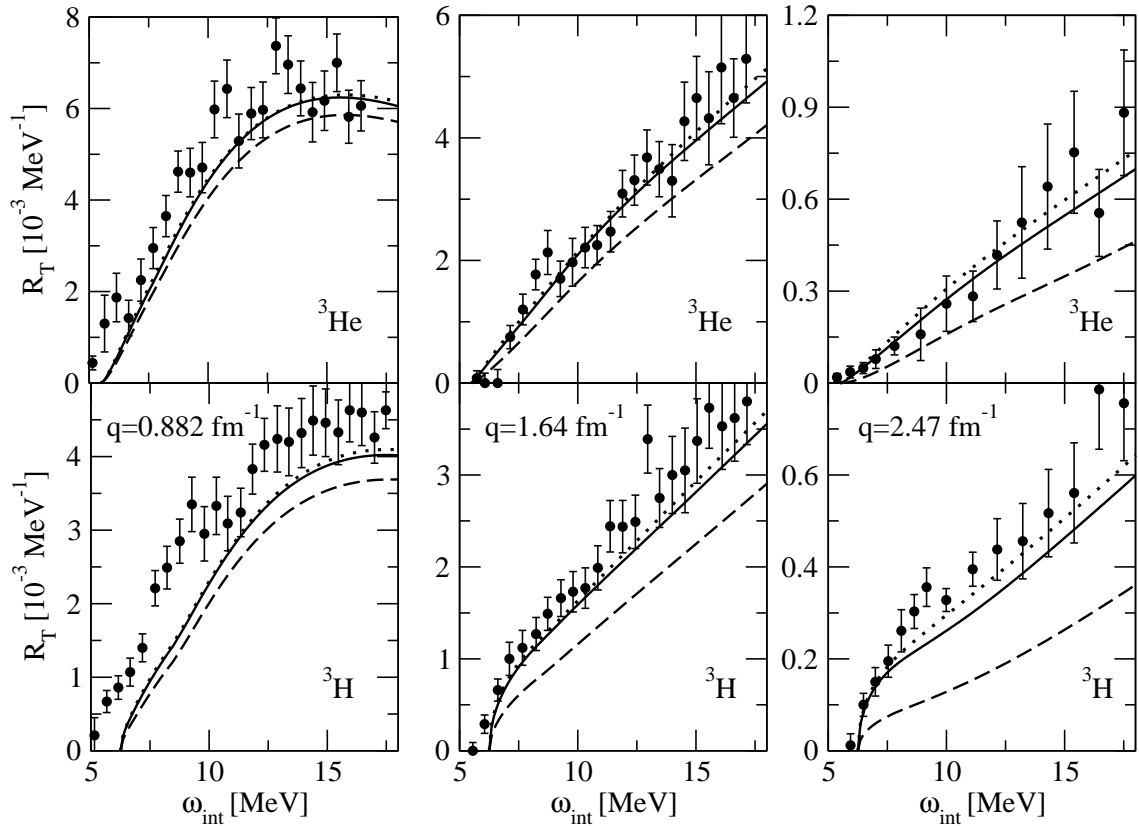}
  \caption{Comparison of experimental and theoretical results for the threshold $^3$He (upper panels)
and $^3$H (lower panels) transverse response function RT as function of internal excitation energy
!int at three momentum transfers q. Theoretical results with different current operators: relativistic one-body
current (dashed), relativistic one-body current + MEC (solid), non-relativistic one-body current +
MEC (dotted). Experimental data from~\cite{Retzlaff}.}
\label{RT_3bodyexc}
\end{figure}

Now I am going to illustrate an instructive example, taken from references~\cite{MEKLOTY} and~\cite{3RTAV18}, of how 
one gets information about the subnuclear d.o.f. from the calculation of $R_T$. 
In~\cite{MEKLOTY} the Bonn potential~\cite{BonnRA}, which is a potential based on meson theory was used, while in~\cite{3RTAV18} 
the  potential was the phenomenological AV18~\cite{AV18}. Therefore in the former case the currents are known, while in the latter
a recipe is proposed in order to construct them so that their divergence is consistent with the potential via the continuity equation.
However, only the comparison with data can say whether they are correct. 
In analogy with 
the one boson exchange potential case these 
currents  are called meson exchange currents and indicated by MEC.
Figure~\ref{RT_3bodyexc} illustrates  the importance of these MEC and the quality of the comparison with existing data. 

In case of $^3$He one has a rather good agreement of theoretical and experimental transverse response 
functions for the two higher $|\vec q|$ values. The MEC contribution is essential for reaching this agreement. 
At $|\vec q|$ = 0.88 fm$^{-1}$, however, the theoretical $R_T$ underestimates the data below 11 MeV. 
In the triton case the situation looks worse. Already for the two higher $|\vec q|$ values one finds a 
slight underestimation of the data, in addition the discrepancy becomes even larger at the lowest $|\vec q|$.
 One can conclude that the present agreement between theory and experiment is not bad, but certainly not very good. 
The interesting point is that it seems that a different nuclear force does not improve the situation, since the  
results at $|\vec q|$ = 0.88 fm$^{-1}$ 
with the BonnA potential from~\cite{MEKLOTY} is almost identical to the AV18 result 
(the $^3$H case was not considered in~\cite{MEKLOTY}). Additional currents involving the $\Delta$ resonance, 
up to now only partially considered in the literature for the threshold kinematics, 
could probably lead to a small improvement.

\subsection{Results for A=4}

Among light nuclei the $\alpha$-particle ($^4$He) is of  particular importance 
because it has some typical features of heavier 
systems (e.g. binding energy per nucleon), which make it a precious link 
between the classical few-body systems, i.e. deuteron, triton  and $^3$He, and more 
complex nuclei. For example in $^4$He one can study the possible emergence of 
collective phenomena typical of complex nuclei like the {\it giant dipole resonance} (GDR).
This is the famous bump which is seen in the total photoabsorption cross section of all nuclei
and that has been interpreted as a collective oscillating motion of the protons against the neutrons.
Furthermore $^4$He is the 
ideal testing ground not only for the NN potential but also for the multi-nucleon forces and in particular 
of the three-body force (3BF).

\subsubsection{Digression on the three-body force}

As was already remarked above, the nuclear potential has clearly an {\it effective} nature, therefore it is in principle 
a many-body operator. Yet the debate has concentrated for several decades only on the two-body part.
Such a debate has taken place essentially among three different points of view: those based on
meson theory, purely phenomenological ones and more recently  
the point of view of  effective field theory. {\it Realistic} potentials based on all three 
different approaches have been constructed, relying on fits to thousands of N-N scattering data. 
After that precise  calculations of the triton binding energy have demonstrated that a two-body 
potential is not enough to explain the experimental value, the same debate is taking place regarding 
three-body forces. However, for the determination of a {\it realistic}  three-body potential or 
to discriminate among different models one needs to find  A$\geq3$ observables that are sensitive 
to it. One direction that has been followed~\cite{wiringa04,wiringa00} is to calculate accurately 
bound properties of nuclei of increasing A. 
In fact it has been  
realized that stronger and stronger discrepancies exist between the binding energies calculated 
with high precision  two-body potentials and the experimental values (see the upper part of figure~\ref{wiringa}). 
Another very promising direction is to study electromagnetic reactions 
in the continuum.
In fact many years of electron scattering experiments have demonstrated the power of 
this kind of reactions, and in particular of the inelastic ones, because of the  
 possibility to vary energy $\omega$ and   
momentum $|\vec q|$ transferred by the electron to the nucleus. This allows one to focus on different dynamical 
aspects at different ranges and one might find regions where the searched three-body effects are sizable. 

The $^4$He nucleus is particularly appropriate for these studies because of the following considerations: 
i) the ratio between the number of triplets and of pairs goes like $(A-2)/3$, therefore it is double for 
$^4$He than for $^3$He; ii) theoretical results on hadron scattering observables involving four 
nucleons~\cite{deltuva} as well as $^4$He-N phase shifst~\cite{sofiascattering} seem to imply that three-body 
effects are rather large.

\subsubsection*{Digression on the experimental situation}

In order to appreciate the importance of the theoretical results presented in the following
it is necessary to summarize the situation of the experiments regarding electromagnetic
reactions on $^4$He.

 Let us start from the real photon case, which has  has a longstanding 
history. 
First experiments on the $(\gamma,n)$ reaction were performed about fifty
years ago.
In the following twentyfive years various experimental 
groups carried out measurements for both the $(\gamma,p)$and 
$(\gamma,n)$ reaction channels and the inverse capture reactions 
Dramatically conflicting results have 
been obtained as a result of this work. The $(\gamma,p)$ data were  
consistent in showing a rather pronounced resonant peak close to the 
three-body breakup threshold. 
At the same time, the $(\gamma,n)$ data at low energy were very spread, 
and measurements showed  either a strongly pronounced  or a rather 
suppressed giant dipole peak. In 1983 a careful and balanced review of 
all the available experimental data for the two mirror reactions was 
provided~\cite{CBD:1983}.
A strongly peaked cross section at low energy  for the ($\gamma,p$) 
channel and a flatter shape for the ($\gamma,n$) one was recommended 
by the authors. Three new experiments on the ($\gamma,p$) reaction 
were subsequently carried out, two of them contradicting and one 
confirming the recommended cross section. Measurements of 
the ratio of the $(\gamma,p)$ to  $(\gamma,n)$  cross sections in the 
giant resonance region were  performed  as well~\cite{FLORIZONE94} and, at variance with 
the cross sections recommended in~\cite{CBD:1983}, results very close to 
unity were reported. Finally, in 1992 additional 
cross section data were deduced from a Compton scattering experiment on 
$^{4}$He~\cite{Wells92}. A strongly peaked cross section for the 
$^{4}$He total photoabsorption was found suggesting a ($\gamma,n$) 
cross section considerably larger  than the one recommended in 
\cite{CBD:1983}. 
 
In  1996 the first theoretical calculation  of the two-fragment breakup 
cross section with inclusion of 
FSI was performed in the energy range below the three-fragment 
breakup threshold~\cite{Ell96}. The semi-realistic MTI-III potential~\cite{MT69} was employed. The results of~\cite{Ell96} showed a rather 
suppressed giant dipole peak. The agreement with the ($\gamma$,n) data in~\cite{Berman} was very good and
the situation seemed to be settled.

However, the following  year a calculation of the 
total  photoabsorption cross section for the $^{4}$He up to the pion threshold
was carried out~\cite{ELO97}. Full FSI was taken into account in 
the whole energy range via the application of the LIT method. The four-nucleon dynamics was 
described with the same NN potential as in~\cite{Ell96}. Different from 
the previous work  a pronounced giant dipole peak was found. 
These results have been reexamined in 
\cite{BELO01,Sofia1}. A small shift of the peak position has been obtained, 
but the pronounced peak has been confirmed. 

Unfortunately, the experimental situation of the $^4$He 
photodisintegration is not yet sufficiently settled. Most of the experimental work has 
concentrated on the two-body break-up channels $^4$He$(\gamma,n)^3$He and
$^4$He$(\gamma,p)^3$H in the giant resonance region, but still today there is large 
disagreement in the peak. In fact in two recent $(\gamma,n)$ experiments 
\cite{Lund,Shima} one finds differences of a factor of two. A measurement of the analog of the GDR in $^4$He
has been performed in~\cite{NAKAYAMA07} via the $^4$He($^7$Li,$^7$Be) and points to a cross section which is 
very close to the result in~\cite{ELO97}. 

Considering that from the early times
of atomic physics the cross
section for absorption of photons has always represented the fundamental observable
to study the spectrum of composite systems, the present situation is unacceptable.
Therefore it is highly desirable that this observable attracts
the interest of experimentalists. (Projects in this direction have been recently considered at MaxLab in Lund 
and HI$\gamma$S at TUNL in North Carolina). 

\bigskip

Regarding the experimental situation for electron scattering, in the '80 and '90's an intense experimental activity has been devoted to 
it, in particular to {\it inclusive} electron 
scattering (denoted by $(e,e')$) in the so called quasi-elastic (q.e.) regime, corresponding to momentum 
transfers of several hundreds MeV/c and energies around the  q.e. peak, observed at about $q^2/2m$. 
In such conditions one can  envisage that the electron has scattered elastically with a single nucleon of 
mass $m$. Various nuclear targets have been considered, from very light 
to heavy ones.
The reason for concentrating on the q.e. regime has been the conviction that for such  kinematics  
the plane wave impulse approximation (PWIA) might 
be a reliable framework to describe the reaction.
The neglect of the final state interaction (FSI) has the advantage to allow a simple interpretation 
of the cross section in terms of the dynamical properties of the nucleons in the ground state. In particular the focus 
in those years were the ground state short range correlations, i.e. the dynamical effects on the wave function of the largely unknown repulsive short range
part of the potential. 

It is clear, however, that it is important to clarify 
the reliability of the PWIA. As it will be seen in the following, the LIT method, which takes into account the full dynamics in the final state, 
allows to clarify this issue.

\subsubsection*{The comparison theory-experiment for the photon case}
 
\begin{figure} 
 \includegraphics[height=.5\textheight]{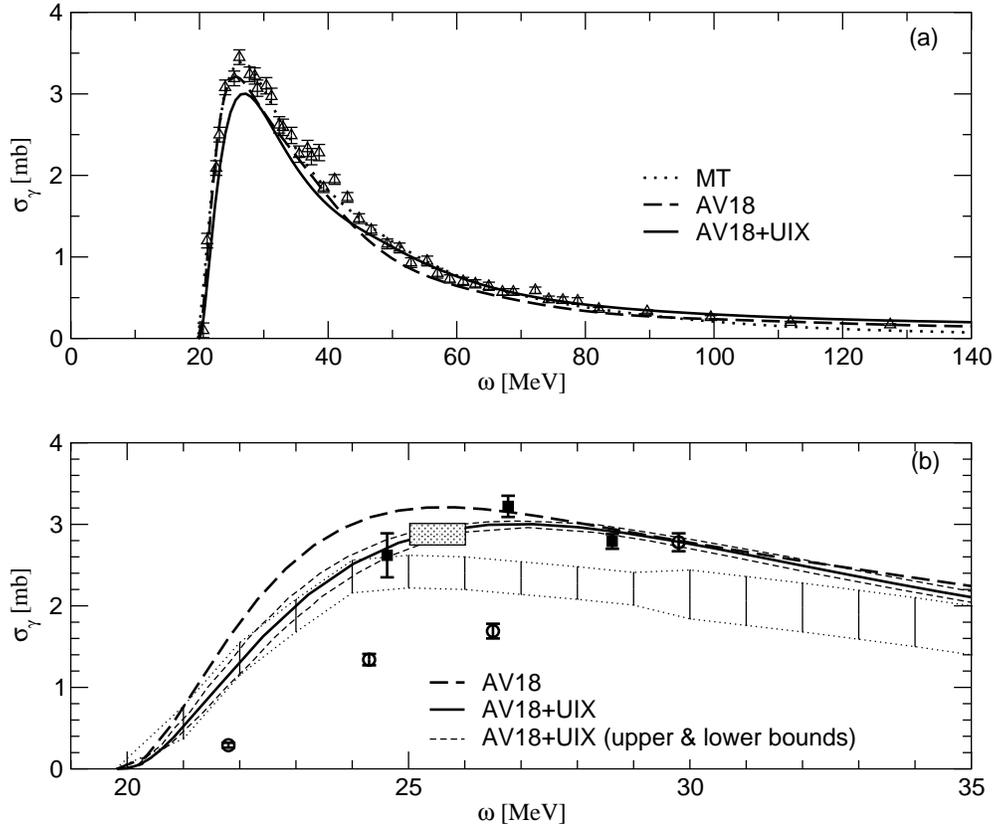}
  \caption{Total $^4$He photoabsorption cross section: (a) $\sigma_\gamma$ (AV18) 
and $\sigma^\infty_{\gamma,19}$ (AV18+UIX),
(b) as (a) but also included upper/lower bounds and various experimental 
data (see text), area between
dotted lines~\cite{Berman,Feldman}, dotted box~\cite{Wells92}, squares 
\cite{Lund}, and circles~\cite{Shima}.}
\label{4photonrealistic}
\end{figure}

An important step forward on the theory side has been made by
performing a calculation of the total photoabsorption cross section $\sigma_\gamma$
of $^4$He with a realistic nuclear force~\cite{PRLPhoton}. To this end the four-body
problem with the Argonne V18 (AV18) NN potential 
\cite{AV18} and the Urbana IX (UIX) 3NF~\cite{AV18+} has been solved.
The results are shown in figure~\ref{4photonrealistic}.

Due to the 3NF one 
observes a reduction of the peak height by about 6\% and a shift of the peak 
position by about 1 MeV towards higher energy. Large effects of the 3NF are 
found above 50 MeV with an increase of $\sigma_\gamma$ by e.g. 18\%, 25\%, and 35\% 
at $\omega=60$, 100, and 140 MeV, respectively. 
It is very interesting to compare the present 3NF effects to those
found for $\sigma_\gamma(^3$H/$^3$He)~\cite{ELOT00,bench3}. Surprisingly, the 
peak height reduction is smaller for $^4$He.
For $^3$H/$^3$He the size of the reduction is similar to the increase of $E_b$ (10\%),
whereas for $^4$He the 3NF increases $E_b$ by 17\%, but reduces the peak
by only 6\% and thus cannot
be interpreted as a simple binding effect. Also at higher $\omega$ there are
important differences. The enhancement of $\sigma_\gamma(^4$He) due to the
3NF is significantly stronger, namely about two times larger than for the
three-body case. Interestingly this reflects the above mentioned different ratios between
triplets and pairs in three- and four-body systems. 
In figure~\ref{4photonrealistic}(a) also $\sigma_\gamma$ for the semi-realistic Malfliet-Tjon (MT) NN
potential~\cite{ELO97,BELO01} is illustrated. Similar to  
$\sigma_\gamma(^3$H/$^3$He)~\cite{ELOT00} one finds a rather "realistic" result in the giant 
resonance region (overestimations of the peak by about 10-15\%) and 
quite a correct result for the peak position; however, at higher energy 
$\sigma_\gamma$ is strongly underestimated, for $^4$He by a factor of three at 
pion threshold. In figure~\ref{4photonrealistic}(a) data from~\cite{Ark} are also shown.
They are the only measurements of $\sigma_\gamma(^4$He) in the whole energy
range up to pion threshold. In the peak region the data agree best with the MT potential,
while for the high-energy tail one finds the best agreement with the AV18 potential.

In figure~\ref{4photonrealistic}(b)  our low-energy results are compared to further data. 
which, however have  to be interpreted with some care, since no of them corresponds to 
a direct measurement of the total photoabsorption cross section: 
(i) in~\cite{Wells92} the peak cross section 
is determined from Compton scattering via dispersion relations, (ii) the dashed 
area corresponds to the sum of cross sections for $(\gamma,n)$ from~\cite{Berman} and 
$(\gamma,p)^3$H from~\cite{Feldman} as already shown in~\cite{ELO97}, 
(iii) the data from the above mentioned recent $(\gamma,n)$ experiment 
\cite{Lund} are included only up to about the three-body break-up threshold, 
where one can rather safely assume that $\sigma_\gamma \simeq 2\sigma(\gamma,n)$ 
(see also~\cite{Sofia1}), (iv) in~\cite{Shima} 
all open channels are considered. One sees that the various data 
are quite different exhibiting maximal deviations of about a factor of 
two. The theoretical $\sigma_\gamma$ agrees quite well with the low-energy 
data of~\cite{Berman,Feldman}. In the peak region, however, the situation is very 
unclear. There is a rather good agreement between the theoretical $\sigma_\gamma$ 
and the data of~\cite{Lund} and~\cite{Wells92}, while those of~\cite{Berman,Feldman} 
are noticeably lower. Very large 
discrepancies  are found in comparison to the 
recent data of Shima et al.~\cite{Shima}, while a very good agreement is found with data in~\cite{NAKAYAMA07}.

From all this long discussion it is evident that the experimental 
situation is rather unsatisfactory and further improvement is urgently needed, if one wants to 
clarify the 3NF issue. In particular it is worth to stress that due to the above mentioned Siegert's theorem
what one is testing here are also three-body exchange currents connected to the 3NF.

\subsubsection*{The comparison theory-experiment for the electron case}

Inelastic electron scattering off nuclei provides complementary informations on the nuclear dynamics.  
In fact varying the momentum $|\vec q|$, transferred by the electron to the nucleus, one
can focus on different  
dynamical regimes. At lower momenta  the collective behavior of nucleons is studied. As  $|\vec q|$  
increases one probes  properties of the  single nucleon in the nuclear medium 
and its correlations to other nucleons from long- to short-range.
Since the longitudinal response $R_L$ is not sensitive to meson exchange
effects (only at lowest relativistic order in the Foldy-Wouthuysen transformation)  
the use of a simple one-body 
density operator allows to concentrate on the nuclear dynamics generated by the potential
and one might find regions where the searched three-nucleon effects are sizable. 

\begin{figure} 
\includegraphics[width=0.7\textwidth]{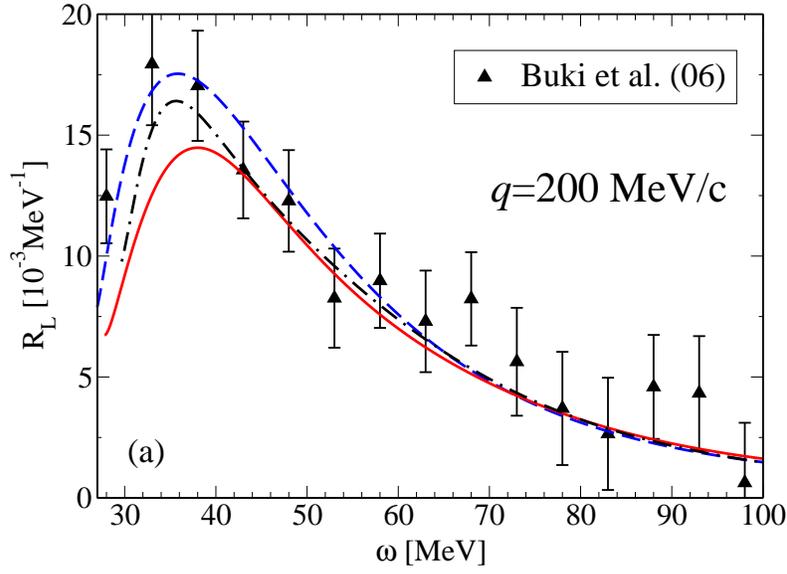}
\caption{Longitudinal response function for $|\vec q|=200$  MeV/c
with the AV18 (dashed), AV18+UIX (solid) and MT (dashed-dotted) potentials. Data from~\cite{Buki06}.} 
\label{electron200_100a}
\end{figure}

\begin{figure} 
\includegraphics[width=0.7\textwidth]{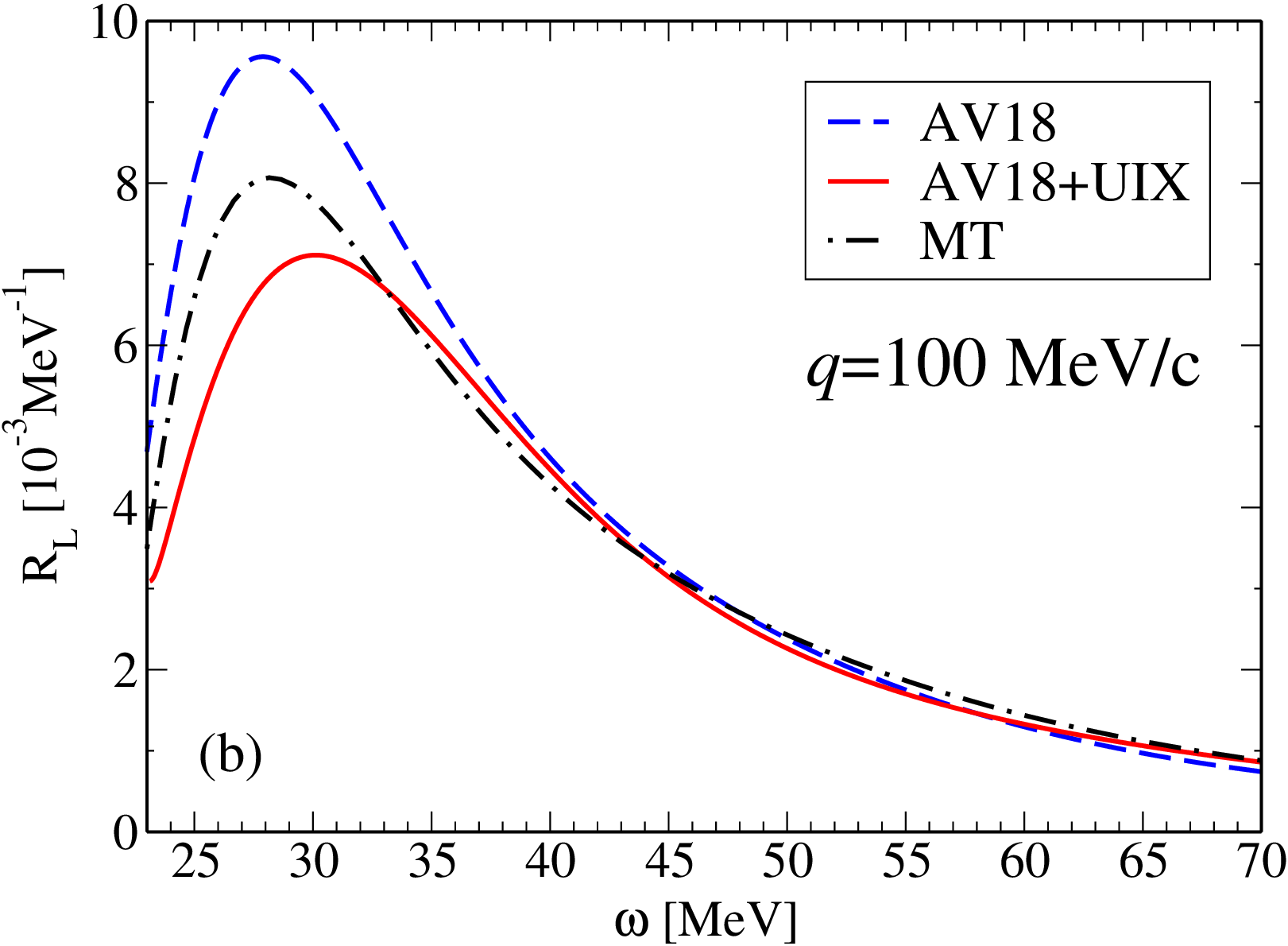}
\caption{Longitudinal response function for  $|\vec q|=100$ MeV/c 
with the AV18 (dashed), AV18+UIX (solid) and MT (dashed-dotted) potentials.}
\label{electron200_100b}
\end{figure}

Here  results on $R_L$ are presented, focusing on the evolution 
of dynamical effects as the momentum transfer decreases. In figures~\ref{electron200_100a} and~\ref{electron200_100b}, 
$R_L$ at constant $|\vec q|= 200$ and 100 MeV/c are shown.
\begin{figure} 
\includegraphics[width=0.7\textwidth]{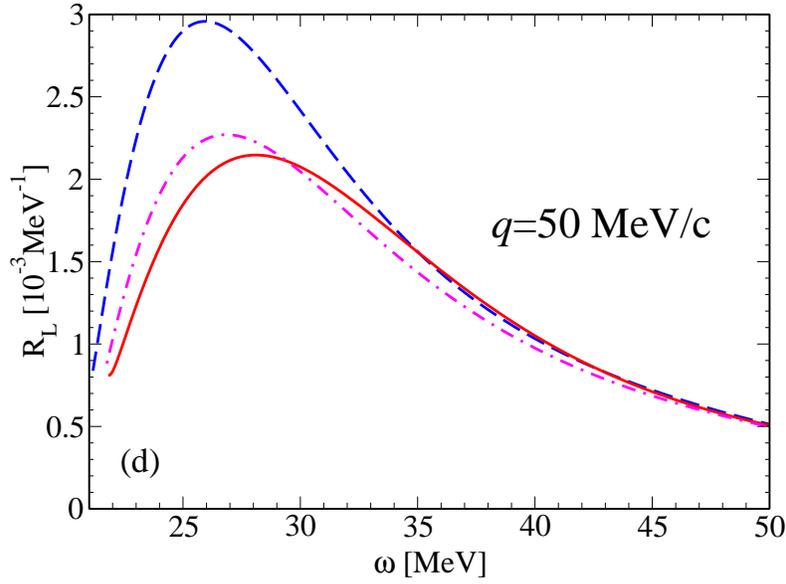}
\caption{Longitudinal response function for  $|\vec q|=100$ MeV/c 
with the AV18 (dashed), AV18+UIX (solid) and MT (dashed-dotted) potentials.}
\label{electron50TMUIX}
\end{figure}
\begin{figure} 
\includegraphics[width=0.9\textwidth]{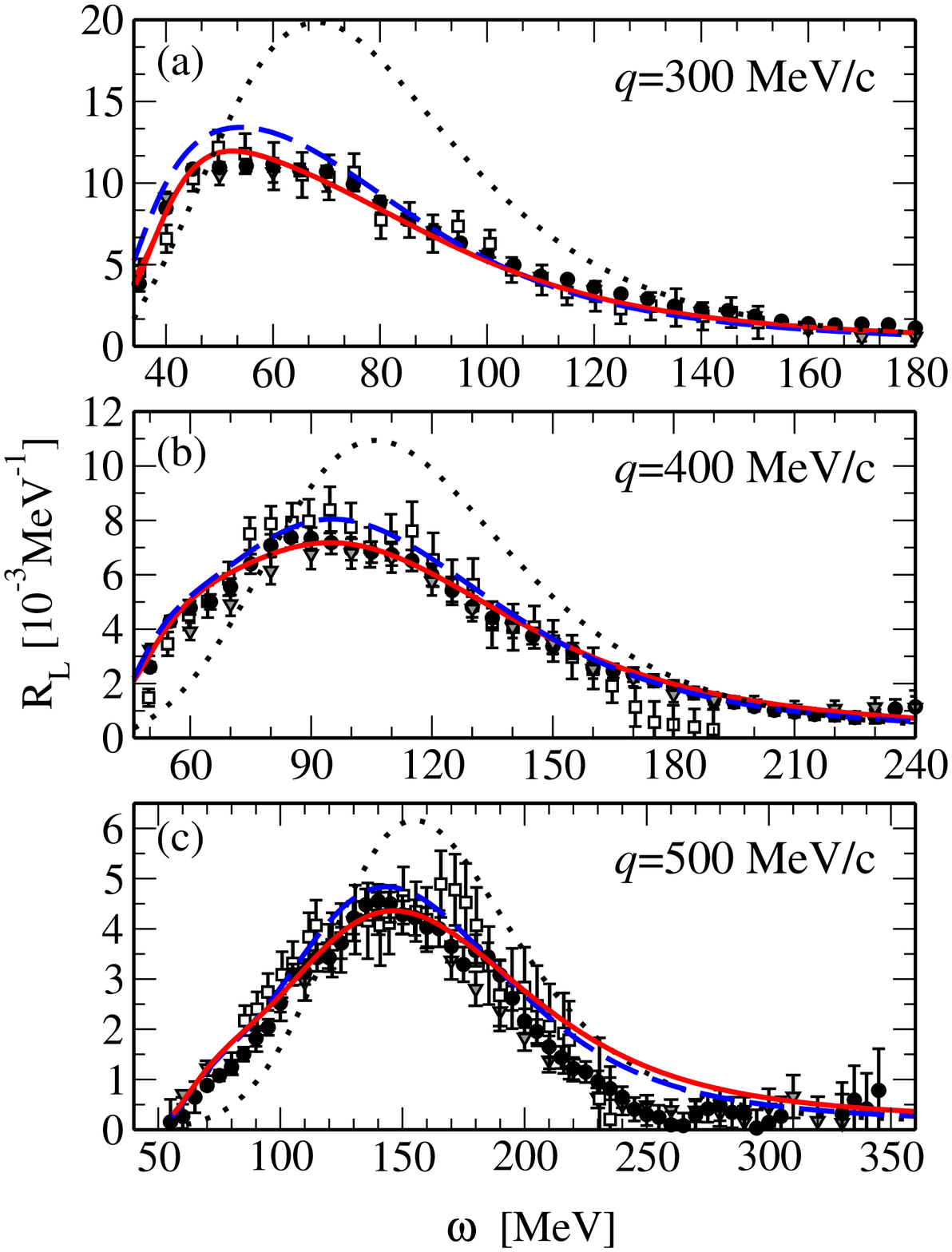}
\caption{$R_L(\omega,q)$ at various $|\vec q|$: PWIA (dotted); full calculation with AV18 (dashed) and AV18+UIX
(solid). Data from Bates~\cite{Bates} (squares), Saclay~\cite{Saclay} (circles) and world-data set from~\cite{SICK} (triangles).}
\label{electronhighq}
\end{figure}
One has a large quenching effect due the 3NF, which 
is strongest at lower  $|\vec q|$. 
One should notice that such an effect is not simply correlated to the under-binding of the AV18
potential. In fact, if this was the case, the results with the Malfliet-Tjon potential 
(MT)~\cite{MT69}, which gives a slight over-binding of $^4$He, would lay even 
below those obtained with AV18+UIX. 
On the contrary the MT curve is situated
between the curves with and without 3NF. 

Given the large 3NF effect at lower $|\vec q|$ it is interesting to see whether
there is a dependence of the results on the  3NF model itself, investigating in addition other
low-q values. To this end
the calculation has been performed using also the Tucson Melbourne (TM')~\cite{TMprime}  three-nucleon force at $|\vec q|$ = 50 MeV/c. 
While the UIX force contains a two-pion exchange and a short range phenomenological term, with two 3NF parameters fitted
on the triton binding energy and on nuclear matter density (in conjunction with the AV18 two-nucleon potential), the TM' force
is not adjusted in this way. It includes two pion exchange terms where the coupling constants are taken from pion-nucleon scattering 
data consistently with chiral symmetry. Figure~\ref{electron50TMUIX} shows that the increase of 3NF effects 
with decreasing $|\vec q|$ is confirmed. Moreover
it becomes evident that also the difference between the results obtained with two  3NF models 
is sizable. One actually finds that
the shift of the peak to higher energies in the case of UIX  generates for $R_L$ a difference up to about 10\%
on the left hand sides of the peaks. This is a very interesting result. It represents the first case of an 
electromagnetic observable considerably dependent on the choice of the 3NF.
In the light of these results it would be very interesting to repeat the calculation with  EFT 
two-and three-body potentials. At the same time it would be highly desirable 
to have precise measurements of $R_L$ at low $|\vec q|$ in order to discriminate between different
nuclear force models.

In figure~\ref{electronhighq} an overview of the results obtained for larger $|\vec q|$ is given, showing also 
the comparison with existing experimental data. One sees that the 3NF results are closer to the data, this 
is particularly evident at $|\vec q|=300$ MeV/c. However, the 3NF effect is generally not as large as for the lower momentum transfers. 
In some cases the quenching of the strength due to the 3NF is  
comparable to the size of the error bars, particularly for the data from~\cite{Bates}.
The largest discrepancies with data are found at $|\vec q|=500$ MeV/c.
While the height of the peak is well reproduced by the result with 3NF, the width
of the experimental peak seems to be somewhat narrower than the theoretical one. On the other hand
 one has to be aware that relativistic effects
are not completely negligible at $|\vec q|=500$ MeV/c. They probably play a similar role as found in the electro-disintegration
of the three-nucleon systems (see e.g.~\cite{RL3B}). In the case of $|\vec q|=$ 250 MeV/c
the experimental results are not sufficiently precise to draw a conclusion.

But the most striking message one gets from those result  is the large FSI effects that one finds
in all cases, an effect that is 
essential for reaching agreement with experiment. 
The PWIA results fail particularly 
in the q.e. peak and at low energies.
In fact while the FSI effect decreases with increasing $|\vec q|$ in the peak region, this is not 
the case at lower energies. This is particularly a bad news if one considers that the high  $|\vec q|$-low $\omega$ region
was considered the best one to focus on ground state short range correlations.

\subsubsection*{Inelastic Neutrino Reactions}

Here I report an example of weak cross section, calculated ab initio by the LIT method, which
serves to clarify a problem of astrophysical interest.
In fact the current theory of core collapse supernova holds some open
questions regarding the explosion mechanism and late stage
nucleosynthesis. In
particular, due to the high abundance of $\alpha$ particles in
the supernova environment, the inelastic neutrino--$^4$He reaction
is of particular relevance. The characteristic temperatures of the
emitted neutrinos are
about $6-10$ MeV for $\nu_{\mu,\tau}$
($\bar{\nu}_{\mu,\tau}$), $5-8$ MeV for $\bar{\nu}_e$, and $3-5$ MeV for
$\nu_e$. 

In~\cite{nirdoronneutrino} a full {\it{ab--initio}} calculation of
the inelastic neutrino--$^4$He reactions has been considered  for the  channels
$^4$He($\nu_x$,$\nu^{\,\prime}_x$)$^4_2$X,
$^4$He($\bar{\nu}_x$,$\bar{\nu}^{\,\prime}_x$)$^4_2$X,
$^4$He($\bar\nu_e$,e$^+$)$^4_1$X, and $^4$He($\nu_e$,e$^-$)$^4_3$X,
 where $x=e, \mu, \tau$ and $^A_Z$X stands for the final state $A$--nucleon
 system, with charge $Z$.

Table
\ref{table3} gives the temperature averaged total neutral
current inelastic cross--section as a function of the neutrino
temperature for the AV8', AV18, and the AV18+UIX nuclear
Hamiltonians and for the AV18+UIX Hamiltonian adding the MEC. From
the table it can be seen that the low--energy cross--section is
rather sensitive to details of the nuclear force model (the effect
of 3NF is about $30\%$). This sensitivity  gradually decreases with
growing energy. In contrast the effect of MEC is rather small, being on the percentage level. 
\begin{table}
\begin{tabular}{c||c|c|c|c}
\hline \hline T [MeV] &  \multicolumn{4}{c} {$\bra \sigma^0_x \ket_T
=  \frac{1}{2} \frac{1}{A} \bra \sigma_{\nu_x}^0+
\sigma_{\overline{\nu}_x}^0 \ket_T$ [$10^{-42}cm^{2}$] }  \\ \hline
 &  AV8' & AV18 & AV18+UIX & AV18+UIX+MEC \\
\hline
 4    &  2.09(-3) & 2.31(-3) & 1.63(-3) & 1.66(-3) \\
 6    &  3.84(-2) & 4.30(-2) & 3.17(-2) & 3.20(-2) \\
 8    &  2.25(-1) & 2.52(-1) & 1.91(-1) & 1.92(-1) \\
 10   &  7.85(-1) & 8.81(-1) & 6.77(-1) & 6.82(-1) \\
 12   &  2.05     & 2.29     & 1.79     & 1.80    \\
 14   &  4.45     & 4.53     & 3.91     & 3.93    \\
\hline \hline
\end{tabular}
\caption{{\label{table3}} Temperature averaged neutral current
inclusive inelastic cross-section per nucleon (in $10^{-42}cm^{2}$)
as a function of neutrino temperature (in MeV). From~\cite{nirdoronneutrino} }
\end{table}

The overall accuracy of these results  is of the order of $5\%$
and mainly due to the strong sensitivity of the
cross--section to the nuclear model. The numerical accuracy of the
calculation is of the order of $1\%$. Considering this, one can say that, 
thanks to the power of the LIT method, an important step has been done
in the path 
towards a more robust and reliable
description of the neutrino heating of the pre--shock region in
core--collapse supernovae, in which $^4$He plays a decisive role.

\subsection{Results for A=6,7}

Increasing the number of nucleons one may hope to find the surge of typical collective effects. This is indeed what happens if 
one study the total photodisintegration cross section of the 6-body nuclei $^6$Li and $^6$He. 
\begin{figure} 
\includegraphics[width=0.7\textwidth]{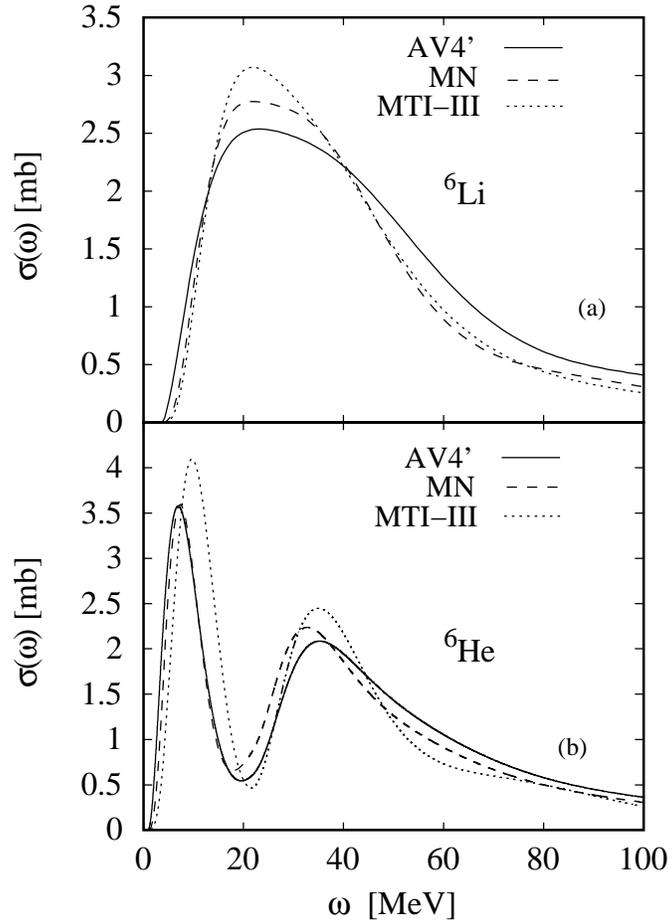}
\caption{Total photoabsorption cross sections for the six-body nuclei with  
AV4',  MN and MTI-III potentials: $^{6}$Li (a), $^{6}$He (b).}
\label{Li6He6}
\end{figure}
\begin{figure} 
\includegraphics[width=0.7\textwidth]{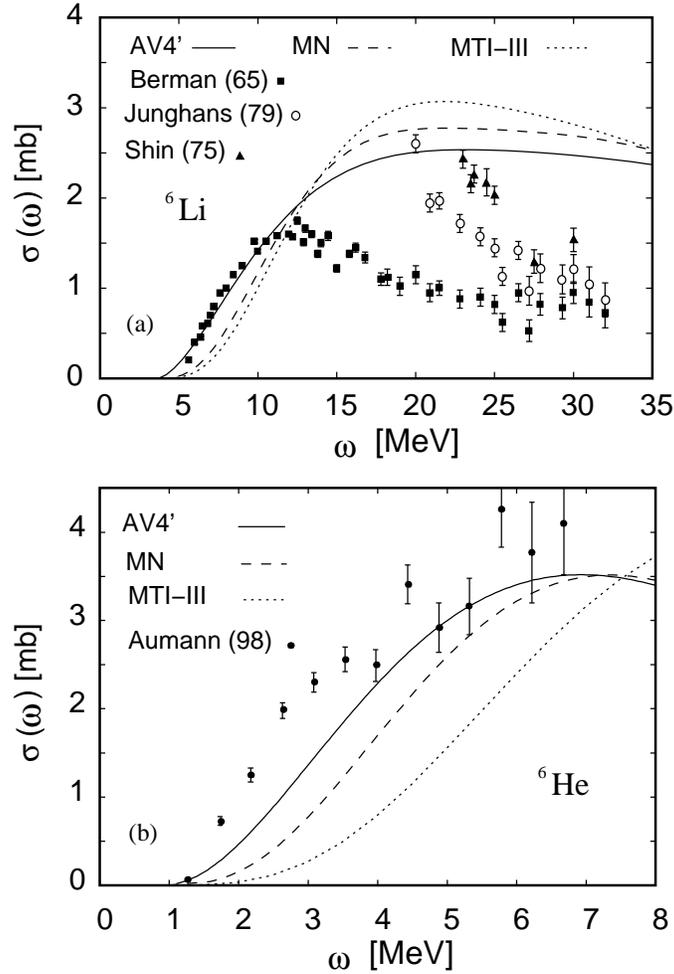}
\caption{Theoretical and experimental photoabsorption cross section results
(see also text).}
\label{Li6He6data}
\end{figure}
\begin{figure} 
\includegraphics[width=0.8\textwidth]{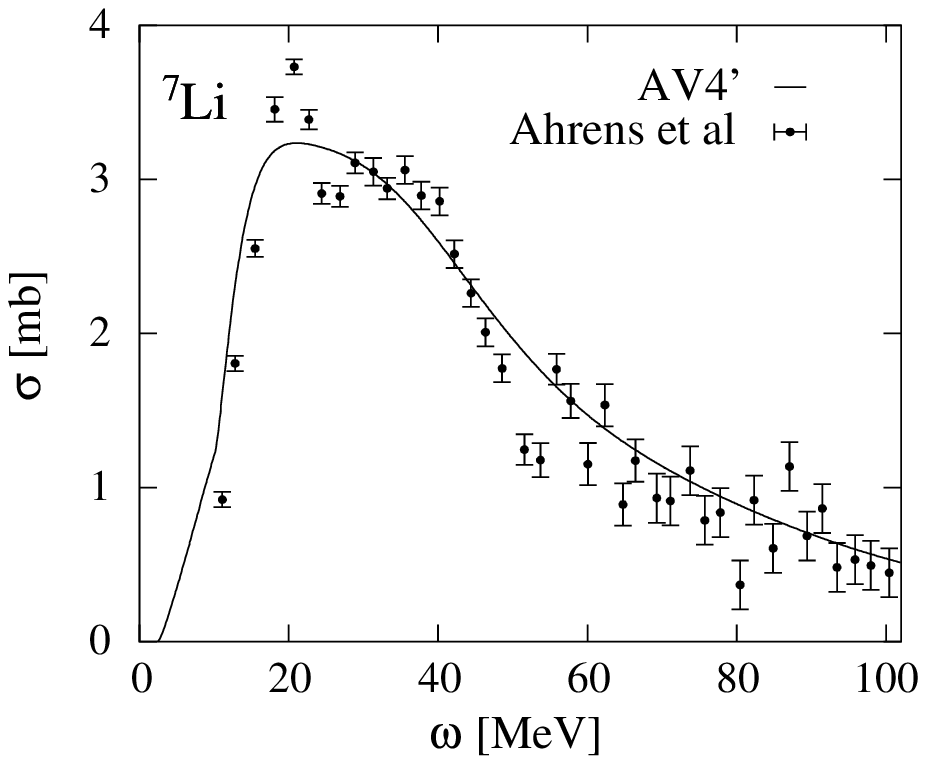}
\caption{Comparison of the theoretical photoabsorption cross section 
calculated with AV4' potential with experimental data from~\cite{Ahr75}.}
\label{Li7}
\end{figure}
In figure~\ref{Li6He6}  the results for the total photoabsorption cross 
section of $^6$Li and  $^6$He~\cite{Li6He6a,Li6He6b} with several semirealistic potentials are shown. 
One notes that the general structure of the cross section is 
similar for the various potential models. In particular one has always the presence of two 
peaks in $^6$He, even if peak positions and peak heights are potential dependent, 
while there is one single giant dipole resonance peak in $^{6}$Li.
The double 
peak structure of $^6$He can be interpreted as a response of a {\it halo} 
nucleus, where the low-energy peak is due to the {\it halo}--$\alpha$ core
oscillation (soft dipole response) and the peak at higher energies due to the 
neutron-proton spheres oscillation (Gamow-Teller mode or hard dipole response).
So the low-energy $^{6}$He peak is due to the
breakup of the neutron halo. The second one, at higher $\omega$, 
corresponds to the breakup of the $\alpha$ core. The 
$^{6}$Li cross section does 
not show such a substructure. This is probably due to the fact that the 
breakup in two three-body nuclei, $^{3}$He~+~$^{3}$H, fills the gap between 
the halo and the $\alpha$ core peaks. Note that in case of $^{6}$He a 
corresponding breakup in two identical nuclei,  $^{3}$H~+~$^{3}$H, is not 
induced by the dipole operator.

In figure~\ref{Li6He6data} the theoretical results are shown together with available 
experimental data. For the AV4' potential~\cite{AV4'} active also in P-wave,
one finds an enhancement of strength 
in the threshold region compared to the S-wave potentials. It is evident
that the inclusion of the P-wave interaction improves the agreement with 
experimental data considerably. This is particularly the case for $^6$Li. In 
fact with the AV4' potential one has a rather good agreement with experimental 
data up to about 12 MeV. In case of $^6$He the increase of low-energy strength 
is not sufficient, there is still some discrepancy with data. Probably, in 
order to describe the {\it halo} structure of this nucleus in more detail 
additional potential parts are needed. In particular the spin-orbit component 
of the NN potential could play a role in the determination of the soft dipole 
resonance. In fact in a single particle picture of $^6$He the two {\it halo} 
neutrons will mainly stay in a p-state and can interact with one of the
core nucleons via the NN LS-force. 

Anyway the experimental situation is very confused. Again, like in the $^4$He case, 
no data comes from a direct measurement of the total photoabsorption cross section.
Such an experiment is now in course at MaxLab in Lund for $^6$Li and we really hope
that they will be able to clarify the situation.

Finally in figure~\ref{Li7} the total photoabsorption cross section of $^7$Li is shown. 
One readily notes that the gross properties of the data, steep rise, 
broad maximum and slow fall off, are very well reproduced quantitatively 
over the whole energy region by the theory. It is worthwhile to 
emphasize that this result is based on an ab initio calculation 
in which the complicated final state interaction of the 7N-system 
is rigorously taken into account by application of the LIT method. 
No adjustable parameters were used, the sole ingredient being the 
AV4' NN potential model. It remains to be seen whether the slight 
variation of the data near and above the maximum will also be found in 
an experiment with improved accuracy. Such an experiment is in course 
at MaxLab in Lund. The new results could clarify in particular the question whether a simple 
semi-realistic potential like the AV4' model is sufficient for 
an accurate theoretical description 
of this reaction or whether a more realistic nuclear force including a 
3N-force is needed. 

\section{conclusions}

In these lectures I have tried to give an overview of what one can learn from studying e.w. 
interactions with nuclei. The essential messages are the following. 
\begin{itemize}
\item
Since the nuclei are interacting strongly and the e.w. interaction is {\it weak} a lowest
order perturbative description of the cross sections allows to focus on the nuclear dynamics. 
\item
Understanding nuclear dynamics means understanding how the {\it effective} 
degrees of freedom that appear in the hamiltonian explicitly (nucleons) interact. Since these d.o.f. are 
not fundamental particles, but composite systems, the interaction is in principle a many-body interaction.
In particular to understand the role of {\it more than two}-body forces in nuclei is an issue which is debated 
at present.
\item
An effective interaction between effective d.o.f.  implies {\it implicit} d.o.f.. They constitute the bridge between
nuclear physics and non perturbative QCD. Electroweak probes, differently from the hadronic ones, are sensitive to them, providing
precious informations about this issue.
\item
In order that a comparison between theory and experiment is meaningful regarding the conclusions one can draw, it is necessary
to work within  theoretical ab initio approaches able to get accurate results with controlled numerical uncertainties.
Due to the difficulties in solving the quantum mechanical many-body problem, especially in the continuum (many particle scattering problem), 
few-body systems assume an important role.
\item
The lorentz integral transform (LIT) method has represented a big step forward in that it has allowed  to calculate ab initio e.w. cross sections
with more than three nucleons in the continuum, reducing the continuum problem to a bound state problem.
\item 
Until now applications of the LIT method, coupled with advanced bound state methods, have concentrated on systems with A=2-7.
Modern realistic potentials, however, have been used up to A=4. The present challenge on the theory side is the extention 
to larger systems, both stable and exotic (halo).
\item 
Confronted with the big progresses of the theory  the experimental situation does not seem to have reached
the same amount of accuracy, hindering in many cases the possibility to draw conclusions from the comparison theory-experiments.
Therefore there is a clear necessity of experimental activity in this field, especially at lower energies and momenta.  
\end{itemize}

In conclusion I hope to have been able to give an idea of the accomplishments in the field of 
e.w. interactions with light nuclear systems, (stable or exotic) and in particular of its future perspectives.
They could be very exciting, considering the rich amount of investigations which are now made possible 
by the  great progress in the field of few-body physics.

\end{document}